\documentclass[aps,prr,reprint,twocolumn,superscriptaddress,floatfix,nofootinbib,longbibliography]{revtex4-1}
\usepackage{epsfig,amsmath,amssymb,color,comment,physics}
\usepackage{bbold}
\usepackage[makeroom]{cancel}
\usepackage[caption=false]{subfig}
\usepackage{mathrsfs}
\usepackage{mathtools}
\usepackage[countmax]{subfloat}
\usepackage[normalem]{ulem}
\usepackage[english]{babel}
\usepackage{dsfont}
\usepackage[bookmarks=true,colorlinks,linkcolor=OrangeRed,urlcolor=NavyBlue,citecolor=RoyalBlue]{hyperref}
\usepackage[dvipsnames]{xcolor}

\usepackage{color}

\definecolor{mypurple}{rgb}{0.49,0.18,0.56}
\definecolor{mygold}{rgb}{0.93,0.59,0.13}
\definecolor{mygreen}{rgb}{0,0.5,0}
\definecolor{myred}{rgb}{0.7,0,0}
\definecolor{myblue}{rgb}{0,0,0.75}
\definecolor{mymagenta}{cmyk}{0,1,0,0.12}
\definecolor{mygray}{rgb}{0.5,0.5,0.5}

\usepackage{umoline}

\definecolor{mypink1}{rgb}{0.858, 0.188, 0.478}

\begin{document}

\title{Stabilizing lattice gauge theories through simplified local pseudogenerators}
\author{Jad C.~Halimeh}
\email{jad.halimeh@physik.lmu.de}
\affiliation{INO-CNR BEC Center and Department of Physics, University of Trento, Via Sommarive 14, I-38123 Trento, Italy}
\author{Lukas Homeier}
\affiliation{Department of Physics and Arnold Sommerfeld Center for Theoretical Physics (ASC), Ludwig-Maximilians-Universit\"at M\"unchen, Theresienstra\ss e 37, D-80333 M\"unchen, Germany}
\affiliation{Munich Center for Quantum Science and Technology (MCQST), Schellingstra\ss e 4, D-80799 M\"unchen, Germany}
\author{Christian Schweizer}
\affiliation{Walther-Mei\ss ner-Institut, Bayerische Akademie der Wissenschaften, D-85748 Garching, Germany}
\affiliation{Fakult\"at f\"ur Physik, Ludwig-Maximilians-Universit\"at M\"unchen, Schellingstra\ss e 4, D-80799 M\"unchen, Germany}
\affiliation{Munich Center for Quantum Science and Technology (MCQST), Schellingstra\ss e 4, D-80799 M\"unchen, Germany}
\author{Monika Aidelsburger}
\affiliation{Fakult\"at f\"ur Physik, Ludwig-Maximilians-Universit\"at M\"unchen, Schellingstra\ss e 4, D-80799 M\"unchen, Germany}
\affiliation{Munich Center for Quantum Science and Technology (MCQST), Schellingstra\ss e 4, D-80799 M\"unchen, Germany}
\author{Philipp Hauke}
\affiliation{INO-CNR BEC Center and Department of Physics, University of Trento, Via Sommarive 14, I-38123 Trento, Italy}
\author{Fabian Grusdt}
\affiliation{Department of Physics and Arnold Sommerfeld Center for Theoretical Physics (ASC), Ludwig-Maximilians-Universit\"at M\"unchen, Theresienstra\ss e 37, D-80333 M\"unchen, Germany}
\affiliation{Munich Center for Quantum Science and Technology (MCQST), Schellingstra\ss e 4, D-80799 M\"unchen, Germany}

\begin{abstract}
The postulate of gauge invariance in nature does not lend itself directly to implementations of lattice gauge theories in modern setups of quantum synthetic matter. Unavoidable gauge-breaking errors in such devices require gauge invariance to be enforced for faithful quantum simulation of gauge-theory physics. This poses major experimental challenges, in large part due to the complexity of the gauge-symmetry generators. Here, we show that gauge invariance can be reliably stabilized by employing simplified \textit{local pseudogenerators} designed such that within the physical sector they act identically to the actual local generator. Dynamically, they give rise to emergent exact gauge theories up to timescales polynomial and even exponential in the protection strength. This obviates the need for implementing often complex multi-body full gauge symmetries, thereby further reducing experimental overhead in physical realizations. We showcase our method in the $\mathbb{Z}_2$ lattice gauge theory, and discuss experimental considerations for its realization in modern ultracold-atom setups.
\end{abstract}

\date{\today}
\maketitle
\tableofcontents

\section{Introduction} Gauge theories are a cornerstone of modern physics \cite{Weinberg_book}, describing the interactions between elementary particles as mediated by gauge bosons. They implement physical laws of nature through local constraints in space and time \cite{Gattringer_book}. A paradigmatic example is Gauss's law in quantum electrodynamics, which enforces an intrinsic relation between the distribution of charged matter and the associated electromagnetic field. 

With the advent of high precision and fine control in quantum synthetic matter (QSM) devices, quantum simulation of lattice gauge theories (LGTs) has become an exciting and promising prospect which may help to overcome the significant challenges in studying LGTs theoretically \cite{Wiese_review,Zohar_review,Dalmonte_review,Pasquans_review,Alexeev_review,aidelsburger2021cold,zohar2021quantum,klco2021standard}. Indeed, recent years have witnessed a surge in experimental efforts to realize gauge theories in such setups \cite{Martinez2016,Muschik2017,Bernien2017,Klco2018,Kokail2019,Goerg2019,Schweizer2019,Mil2020,Klco2020,Yang2020,Zhou2021}. Though a postulate in nature, gauge symmetry must be engineered in QSM devices with both matter and gauge fields. This poses a major challenge given the plethora of local constraints that need to be controlled. Various methods have been proposed to stabilize gauge invariance in QSM implementations, with the most popular being those based on energy-penalty schemes \cite{Zohar2011,Zohar2012,Banerjee2012,Zohar2013,Hauke2013,Stannigel2014,Kuehn2014,Kuno2015,Yang2016,Kuno2017,Negretti2017,Dutta2017,Barros2019,Lamm2020,Halimeh2020e,Kasper2021nonabelian,Halimeh2021gauge}. Despite recent progress \cite{Yang2020}, such schemes still require significant experimental overhead, since they involve terms quadratic or, at best, linear in complex often multi-body gauge-symmetry generators \cite{Halimeh2020e}.

\begin{figure}[t!]
	\centering
	\includegraphics[width=.48\textwidth]{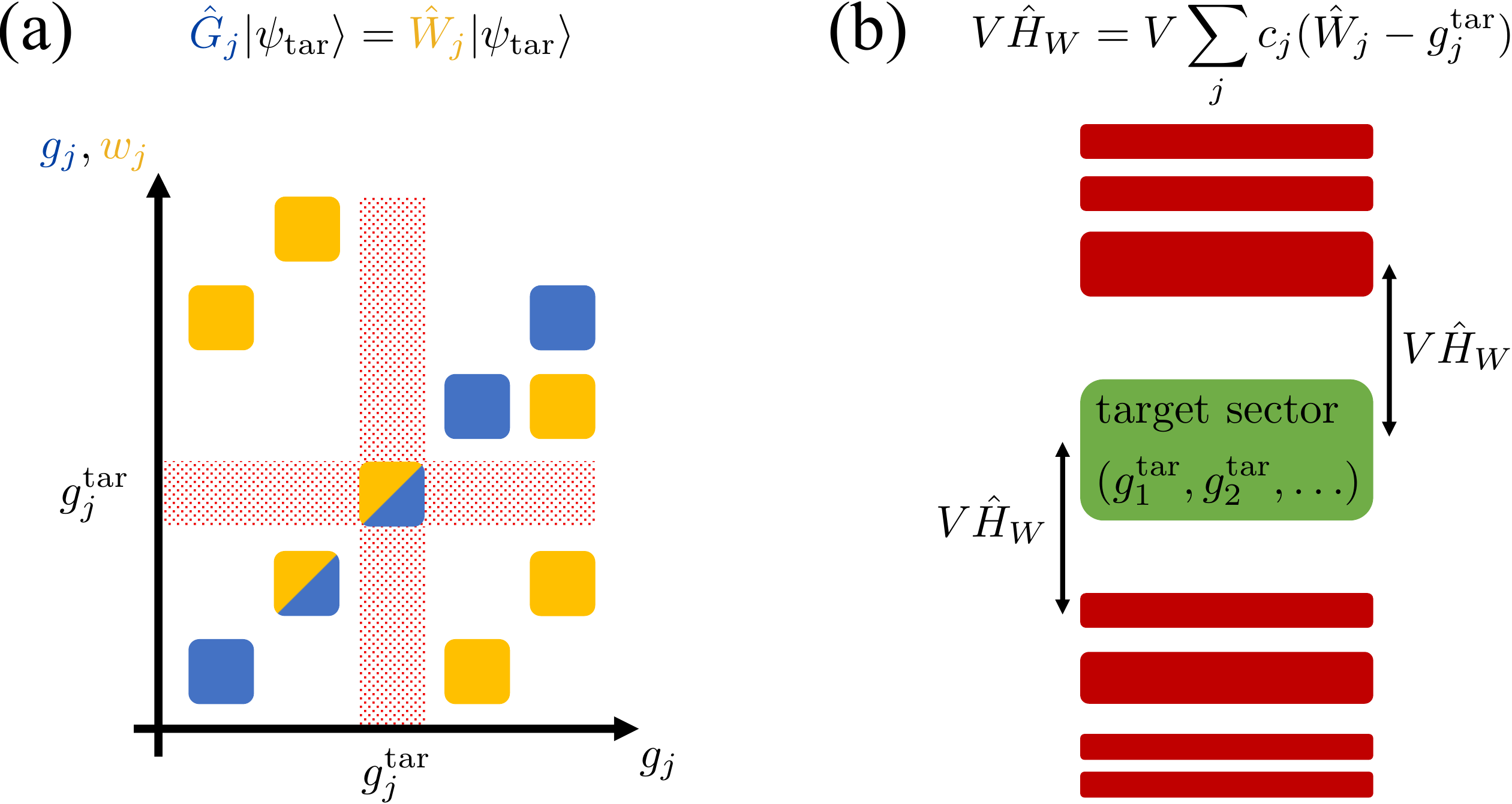}
	\caption{(Color online). Schematic of gauge protection based on the local pseudogenerator (LPG) $\hat{W}_j$ with eigenvalues $w_j$ (yellow boxes), defining a local constraint $j$. (a) When the full generator $\hat{G}_j$ (blue boxes) has an eigenvalue $g_j=g_j^\text{tar}$ in the target sector, then $w_j=g_j^\text{tar}$ and vice versa (yellow/blue box in the middle). When $g_j\neq g_j^\text{tar}$, $w_j$ can have one or more values, one of which may be equal to $g_j$ (yellow/blue box on the left), but never $g_j^\text{tar}$ (forbidden red-dotted regions), or $w_j$ can have no values in the most general case $[\hat{G}_j,\hat{W}_j]\neq0$. (b) In the presence of gauge-breaking errors at strength $\lambda$, the target sector $(g_1^\text{tar},g_2^\text{tar},\ldots)$ is energetically isolated by the LPG protection $V\hat{H}_W$, where $c_j\in\mathds{R};\,\sum_jc_j[w_j(g_j^\text{tar})-g_j^\text{tar}]=0\iff w_j(g_j^\text{tar})=g_j^\text{tar},\,\forall j$. At sufficient strength $V$, LPG protection induces an emergent global symmetry that coincides with the local gauge symmetry within the target sector.}
	\label{fig:schematic} 
\end{figure}

We introduce the concept of the local pseudogenerator (LPG), which is designed to behave identically to the full generator within, but not necessarily outside, the target sector; see Fig.~\ref{fig:schematic}. This relieves significant engineering requirements, rendering the LPG with fewer-body terms than its full counterpart. As we demonstrate numerically and analytically, this approach is powerful---suppressing even nonlocal errors up to all accessible times---and the LPG is readily implementable in modern quantum-simulation platforms, e.g., ultracold atoms and superconducting qubits.

The rest of our paper is structured as follows: In Sec.~\ref{sec:LPG}, we outline the concept and theory of local pseudogenerators. We demonstrate the efficacy of LPG gauge protection in the $(1+1)-$D and $(2+1)-$D $\mathbb{Z}_2$ lattice gauge theory in Secs.~\ref{sec:1DZ2LGT} and~\ref{sec:2D}, respectively. We summarize our results and provide an outlook in Sec.~\ref{sec:summary}. Appendix~\ref{sec:supporting} contains supporting numerical results and Appendix~\ref{sec:analytics} includes our detailed analytic derivations.

\section{Local-pseudogenerator gauge protection}\label{sec:LPG}
In an LGT, couplings between matter and gauge fields have to follow a certain set of rules dictated by the generators of gauge symmetry $\hat{G}_j$ in order to fulfill Gauss's law. Here, $j$ denotes the sites of the lattice, where the matter fields are located, the gauge fields live on links in between sites, and we consider Abelian gauge symmetries. Gauge invariance is embodied in the conservation of all $\hat{G}_j$ by the system Hamiltonian $\hat{H}_0$: $[\hat{H}_0,\hat{G}_j]=0,\,\forall j$. This leads to physical sectors which are characterized by conserved quantum numbers given by the eigenvalues $g_j$ of $\hat{G}_j$. These in turn specify the allowed distributions of matter and the corresponding configurations of electric flux. We denote the desired target sector as the set of all states $\{\ket{\psi_\text{tar}}\}$ satisfying $\hat{G}_j\ket{\psi_\text{tar}}=g_j^\text{tar}\ket{\psi_\text{tar}},\,\forall j$.

The implementation of $\hat{H}_0$ in a realistic QSM setup will lead to gauge-breaking errors $\lambda \hat{H}_\text{err}$ at strength $\lambda$, which couple sectors with different $g_j$. These can be reliably suppressed using the energy-penalty term $V\hat{H}^\text{pen}_G=V\sum_j(\hat{G}_j-g_j^\text{tar})^2$ at sufficiently large positive protection strength $V$ \cite{Halimeh2020a}. Effectively then, $V\hat{H}^\text{pen}_G$ brings the target sector within the ground-state manifold, and any processes driving the system away from it are rendered energetically unfavorable.

Generically, $V\hat{H}^\text{pen}_G$ is experimentally very challenging to realize. Recently, however, protection terms linear in $\hat{G}_j$ have been proposed in the form of $V\hat{H}^\text{lin}_G=V\sum_jc_j(\hat{G}_j-g_j^\text{tar})$ \cite{Halimeh2020e}. If the coefficients $c_j$ are real numbers such that $\sum_jc_j(g_j-g_j^\text{tar})=0$ if and only if $g_j=g_j^\text{tar},\,\forall j$, then gauge invariance can be reliably stabilized up to all accessible times \cite{Halimeh2020e}. Such a sequence $c_j$ has been referred to as \textit{compliant}. Using such \textit{linear gauge protection} may mean the difference between implementing quartic or quadratic terms, such as in the case of $\mathrm{U}(1)$ LGTs \cite{Halimeh2020e}. However, in the case of other models, such as $\mathbb{Z}_2$ LGTs, $(\hat{G}_j-g_j^\text{tar})^2\propto \hat{G}_j-g_j^\text{tar}$, with $\hat{G}_j$ composed of complex multi-body multi-species terms \cite{Borla2019}. In this case, linear protection offers no advantage over its quadratic energy-penalty counterpart.

The major contribution of this work is to introduce the concept of \textit{local pseudogenerators} $\hat{W}_j(g_j^\text{tar})$, see Fig.~\ref{fig:schematic}, which must satisfy the relation
\begin{align}\label{eq:LPGdef}
\hat{G}_j\ket{\psi}=g_j^\text{tar}\ket{\psi}\iff \hat{W}_j(g_j^\text{tar})\ket{\psi}=g_j^\text{tar}\ket{\psi}.
\end{align}
Note that $\hat{W}_j(g_j^\text{tar})$ is dependent on $g_j^\text{tar}$ and is required to act identically to $\hat{G}_j$ \textit{only within} the local target sector, but not necessarily outside it. Indeed, $\hat{W}_j(g_j^\text{tar})$ and $\hat{G}_j$ do not need to commute. This naturally relaxes the engineering overhead on $\hat{W}_j(g_j^\text{tar})$, reducing its number of interacting particles per term relative to $\hat{G}_j$. This technical advantage is the main motivation behind the concept of LPGs. One can now employ the principle of linear gauge protection \cite{Halimeh2020e} using the LPG, rather than the full generator $\hat{G}_j$, through the term
\begin{align}\label{eq:LPGprotection}
V\hat{H}_W=V\sum_jc_j\big[\hat{W}_j(g_j^\text{tar})-g_j^\text{tar}\big],	
\end{align}
which ensures reliably suppression of violations due to any coherent local gauge-breaking errors when the condition $\sum_jc_j[w_j(g_j^\text{tar})-g_j^\text{tar}]=0\iff w_j(g_j^\text{tar})=g_j^\text{tar},\,\forall j$, is satisfied (i.e., $c_j$ is compliant), where $w_j(g_j^\text{tar})$ is the eigenvalue of $\hat{W}_j(g_j^\text{tar})$. Nevertheless, as we will demonstrate in the following, a noncompliant sequence can still reliably stabilize gauge invariance in the case of local gauge-breaking errors up to all accessible times.

\section{$(1+1)-$D $\mathbb{Z}_2$ lattice gauge theory}\label{sec:1DZ2LGT}
Inspired by a recent ultracold-atom implementation \cite{Barbiero2019,Schweizer2019}, we consider the $\mathbb{Z}_2$ LGT in $(1+1)-$D described by the Hamiltonian \cite{Zohar2017,Borla2019,Yang2020fragmentation,kebric2021confinement}
\begin{align}
	\hat{H}_0=J\sum_{j=1}^{L-1}\big(\hat{a}^\dagger_j\hat{\tau}^z_{j,j+1}\hat{a}_{j+1}+\text{H.c.}\big)-h\sum_{j=1}^L\hat{\tau}^x_{j,j+1},
\end{align}
where the local generator of gauge invariance is
\begin{align}\label{eq:Gj}
	\hat{G}_j=(-1)^{\hat{n}_j}\hat{\tau}^x_{j-1,j}\hat{\tau}^x_{j,j+1},
\end{align}
with eigenvalues $g_j=\pm1$. The Pauli matrices $\hat{\tau}_{j,j+1}^{x,z}$ denote the electric and gauge fields, respectively, on the link between matter sites $j$ and $j+1$, and $\hat{a}_j,\hat{a}_j^\dagger$ are hard-core bosonic ladder operators on matter site $j$, with $\hat{n}_j=\hat{a}_j^\dagger \hat{a}_j$ the bosonic number operator. As $\hat{H}_0$ is gauge-invariant, it satisfies $[\hat{H}_0,\hat{G}_j]=0,\,\forall j$.

\begin{table}[t!]
	\centering
	\begin{tabular}{||c | c | c || c | c | c ||} 
		\hline
		 $\hat{n}_j$ & $\hat{\tau}_{j-1,j}^x$ & $\hat{\tau}_{j,j+1}^x$ & $\hat{G}_j$ & $\hat{W}_j(g_j^\text{tar}={\color{myred}-1})$ & $\hat{W}_j(g_j^\text{tar}= {\color{mygreen}+1})$ \\ [0.5ex] 
		\hline\hline
		$0$ &$-1$ &  $-1$ & $\color{mygreen}+1$ & $+1$ & \color{mygreen}$+1$\\ 
		\hline
		$0$ & $-1$ & $+1$ & \color{myred}$-1$  & \color{myred}$-1$ & $-1$\\
		\hline
		$0$ & $+1$  & $-1$ & \color{myred}$-1$ & \color{myred}$-1$ & $-1$ \\
		\hline
		$0$ & $+1$ & $+1$ & \color{mygreen}$+1$ & $+1$ & \color{mygreen}$+1$\\
		\hline
		$1$ & $-1$ & $-1$ & \color{myred}$-1$ & \color{myred}$-1$ & $+3$ \\
		\hline
		$1$ & $-1$ & $+1$ & \color{mygreen}$+1$ & $-3$ & \color{mygreen}$+1$\\
		\hline
		$1$ & $+1$ & $-1$ & \color{mygreen}$+1$ & $-3$ & \color{mygreen}$+1$\\
		\hline
		$1$ & $+1$ & $+1$ & \color{myred}$-1$ & \color{myred}$-1$ & $+3$\\ [1ex] 
		\hline
	\end{tabular}
	\caption{Eigenvalues $g_j$ and $w_j$ of the local full generator $\hat{G}_j$ and the local pseudogenerator $\hat{W}_j$, respectively, for the different possible configurations of the fields on the local constraint specified by matter site $j$ and its neighboring links. Whenever either generator has an eigenvalue $g_j^\text{tar}$, the other does too, i.e., $g_j=g_j^\text{tar}\iff w_j=g_j^\text{tar}$. Contrapositively, whenever either is not $g_j^\text{tar}$, neither is the other: $g_j\neq g_j^\text{tar}\iff w_j\neq g_j^\text{tar}$, though $w_j$ and $g_j$ need not be equal in this case. In our numerical simulations, we have chosen the target sector to be $g_j^\text{tar}=+1$ (green entries), but the conclusions are unaltered for $g_j^\text{tar}=-1$ (red entries), as our method is general and independent of the particular choice of the local target sector.}
	\label{Table1D}
\end{table}

Following the prescription of the LPG given in Eq.~\eqref{eq:LPGdef}, a suitable LPG for $\hat{G}_j$ of Eq.~\eqref{eq:Gj} is
\begin{align}\label{eq:pseudoGen}
	\hat{W}_j(g_j^\text{tar})=\hat{\tau}^x_{j-1,j}\hat{\tau}^x_{j,j+1}+2g_j^\text{tar}\hat{n}_j,
\end{align}
We find that $\hat{W}_j(g_j^\text{tar})\ket{\psi}=g_j^\text{tar}\ket{\psi},\,\forall j$, if and only if $\ket{\psi}$ is in the target sector; see Table~\ref{Table1D}. We emphasize that $\hat{W}_j$ is not an actual local generator of the $\mathbb{Z}_2$ gauge symmetry. In fact, $[\hat{H}_0,\hat{W}_j]\neq0,\,\forall j$.

In the following, we will numerically test gauge protection based on the LPG. Without loss of generality, we will henceforth select the target gauge sector to be $g_j^\text{tar}=+1,\,\forall j$.

\begin{figure}[t!]
	\centering
	\includegraphics[width=.48\textwidth]{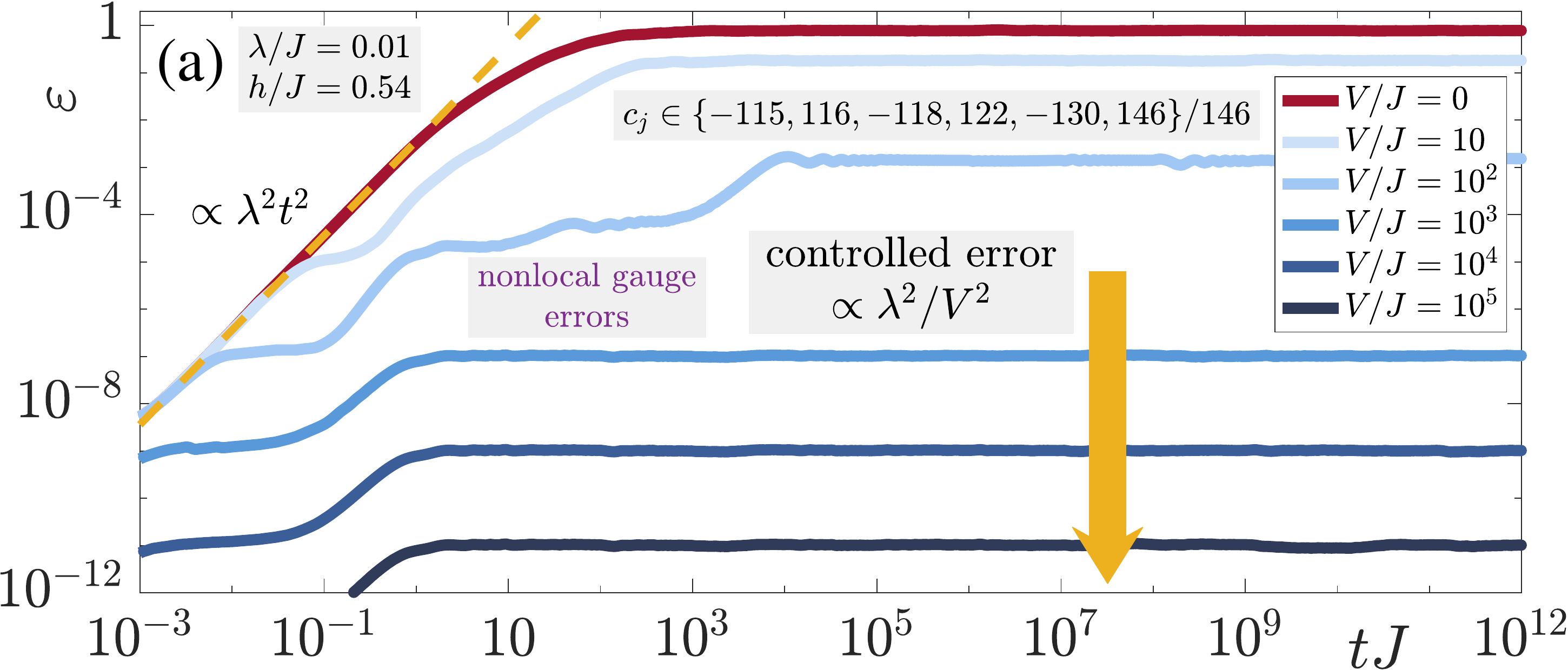}\\
	\vspace{1.1mm}
	\includegraphics[width=.48\textwidth]{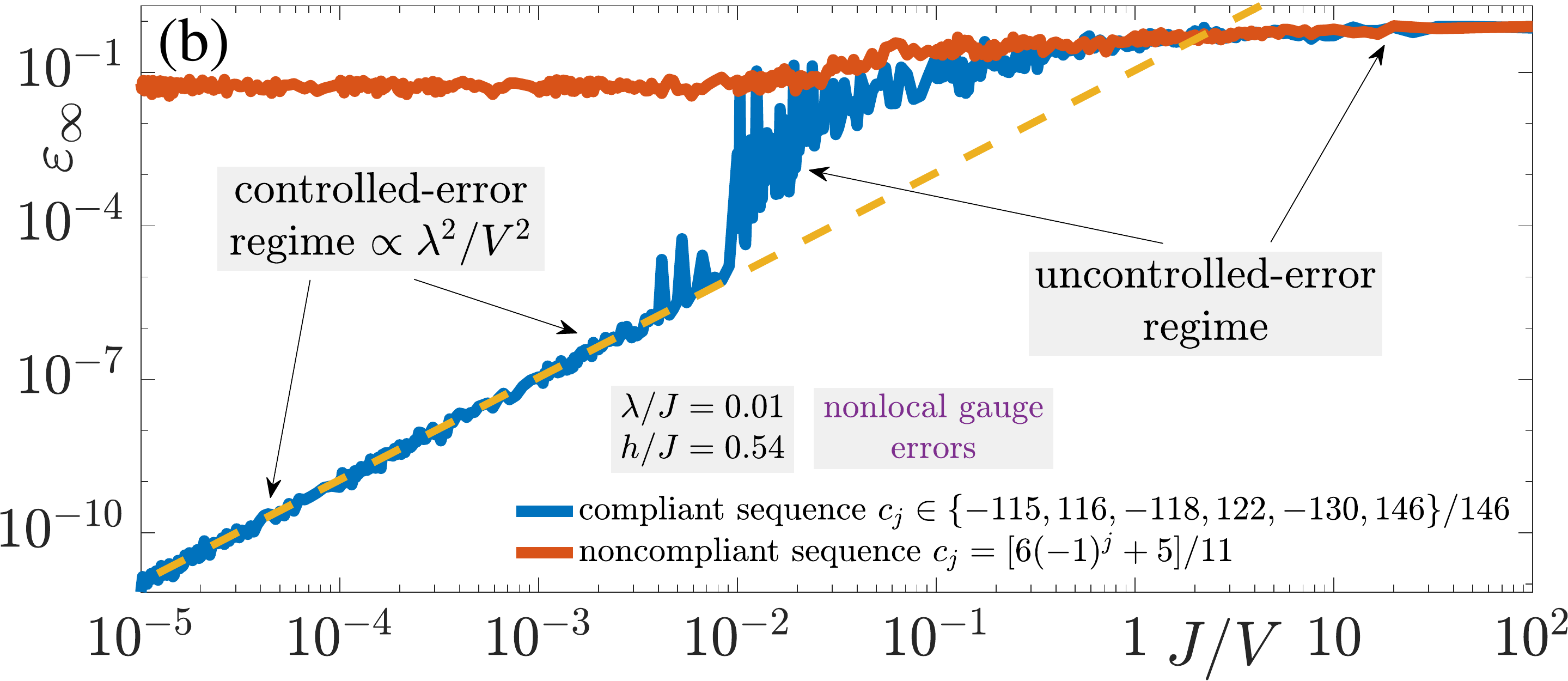}\\
	\vspace{1.1mm}
	\includegraphics[width=.48\textwidth]{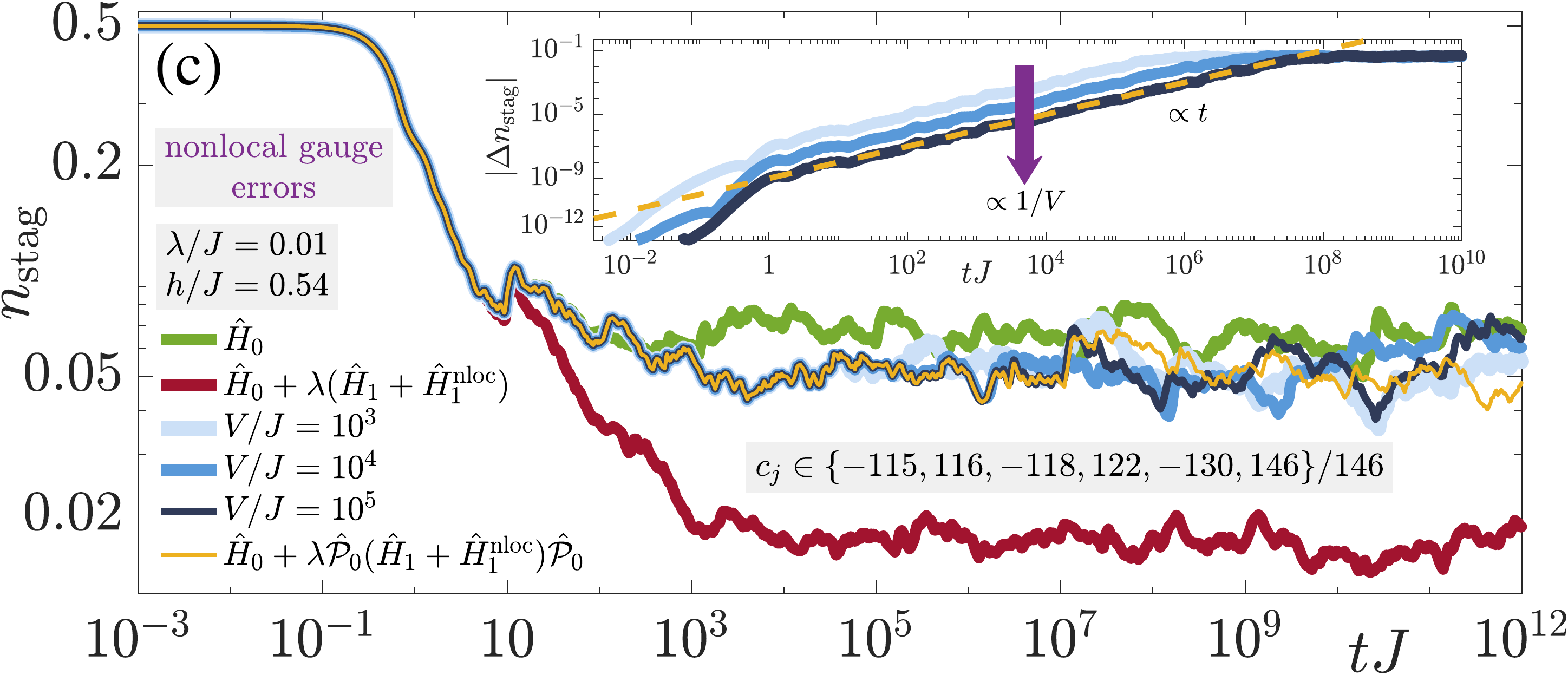}
	\caption{(Color online). Quench dynamics of an initial state in the target sector under the faulty gauge theory $\hat{H}=\hat{H}_0+\lambda(\hat{H}_1+\hat{H}_1^\text{nloc})+V\hat{H}_W$ with experimentally relevant local coherent gauge-breaking errors~\eqref{eq:H1} in addition to the nonlocal error term~\eqref{eq:nonloc}. Protection term~\eqref{eq:LPGprotection} is based on the local pseudogenerator $\hat{W}_j$ given in Eq.~\eqref{eq:pseudoGen} with a compliant sequence. Results are obtained from ED. (a) Gauge violation~\eqref{eq:viol} for various values of protection strength $V$ at an error strength of $\lambda/J=0.01$ with $h/J=0.54$. At sufficiently large $V$, the gauge violation settles at a timescale $\propto1/V$ into a steady state of value $\propto\lambda^2/V^2$. (b) The ``infinite-time'' gauge violation ($t=10^{5}/J$ or larger in ED) shows two distinct behaviors for the compliant sequence. At sufficiently large (small) $V$, it enters a controlled (uncontrolled) error regime where it scales $\propto\lambda^2/V^2$ (displays chaotic behavior). Noncompliant sequences fail to control nonlocal errors. (c) Staggered magnetization shown for the ideal theory (green), under the faulty gauge theory at zero protection strength (red) and several finite values of $V$ (shades of blue), and under the adjusted gauge theory $\hat{H}_\text{adj}=\hat{H}_0+\lambda\hat{\mathcal{P}}_0(\hat{H}_1+\hat{H}_1^\text{nloc})\hat{\mathcal{P}}_0$. At sufficiently large $V$, the dynamics under $\hat{H}$ is reproduced by $\hat{H}_\text{adj}$ within an error $\propto tV_0^2L^2/V$, i.e., up to a timescale $\tau_\text{adj}\propto V/(V_0L)^2$, with $V_0$ an energy scale dependent on the model parameters (but not $V$), as we analytically predict (see Appendix~\ref{sec:analytics}).}
	\label{fig:nonlocal} 
\end{figure}

\subsection{Local and nonlocal gauge errors} We prepare our system in the staggered-matter initial state $\ket{\psi_0}$ in the target sector (see Appendix~\ref{sec:supporting} for details), and quench it with the \textit{faulty} gauge theory
$\hat{H}=\hat{H}_0+\lambda(\hat{H}_1+\hat{H}_1^\text{nloc})+V\hat{H}_W$, where 
\begin{align}\nonumber
	\hat{H}_1=\,\sum_{j=1}^{L-1}\Big[&\big(\alpha_1\hat{a}_j^\dagger\hat{\tau}^+_{j,j+1} \hat{a}_{j+1}+\alpha_2\hat{a}_j^\dagger \hat{\tau}^-_{j,j+1} \hat{a}_{j+1}+\mathrm{H.c.}\big)\\\label{eq:H1}
	&+\big(\alpha_3\hat{n}_j-\alpha_4\hat{n}_{j+1}\big)\hat{\tau}^z_{j,j+1}\Big],
\end{align}
is an experimentally relevant local error term inspired from the setup of Ref.~\cite{Schweizer2019}. The coefficients $\alpha_{1,\ldots,4}$ are real numbers whose relative values depend on the driving parameter in the Floquet setup used to implement $\hat{H}_0$; cf.~Appendix~\ref{sec:coeffs} for exact expressions. Here, we normalize them such that their sum is unity in order to encapsulate the error strength in $\lambda$. We additionally include the nonlocal error term
\begin{align}\label{eq:nonloc}
	\hat{H}_1^\text{nloc}=\sum_{\xi=\pm1}\prod_{j=1}^{L}\big(\mathds{1}+\xi\hat{\tau}^z_{j,j+1}\big),
\end{align}
which though very unlikely to occur in typical experimental setups, is ideal to scrutinize the efficacy of the LPG protection. Note that $\hat{H}_0$, $\hat{H}_1$, and $\hat{H}_1^\text{nloc}$ all conserve boson number, which allows us to work within a given sector of the corresponding global $\mathrm{U}(1)$ symmetry. This permits in exact diagonalization (ED) system sizes of $L=6$ matter sites and $L=6$ gauge links (equivalent to $12$ spin-1/2 degrees of freedom) in the bosonic half-filling sector. However, our method also works for errors violating both the global $\mathrm{U}(1)$ symmetry and the local $\mathbb{Z}_2$ gauge symmetry, and also for different initial states and model-parameter values (see Appendix~\ref{sec:supporting} for supporting results). We employ open boundary conditions for experimental relevance.

Suppression of gauge violations due to gauge-breaking terms such as those of Eqs.~\eqref{eq:H1} and~\eqref{eq:nonloc} has been shown to be effective using the ``full'' protection term $V\hat{H}_G^\text{pen}=V\sum_j(\hat{G}_j-g_j^\text{tar})^2=2V\sum_jg_j^\text{tar}(g_j^\text{tar}-\hat{G}_j)$ in the $(1+1)-$D $\mathbb{Z}_2$ LGT \cite{Halimeh2020a}. This term is complicated to implement experimentally owing to $\hat{G}_j$ containing three-body terms; cf.~Eq.~\eqref{eq:Gj}. This is the main reason why the LPG protection~\eqref{eq:LPGprotection} is ideal here, given that $\hat{W}_j$ includes single and two-body terms only; see Eq.~\eqref{eq:pseudoGen}. Indeed, the level of difficulty for implementing $\hat{W}_j$ is lower than that of the ideal gauge theory $\hat{H}_0$ itself.

We are interested in the dynamics of local observables in the wake of the quench. In particular, we analyze the temporally averaged gauge violation and staggered boson number
\begin{subequations}
\begin{align}\label{eq:viol}
		\varepsilon(t)&=1-\frac{1}{Lt}\int_0^t ds\,\sum_{j=1}^L\bra{\psi(s)}\hat{G}_j\ket{\psi(s)},\\\label{eq:nstag}
		\hat{n}_\text{stag}(t)&=\frac{1}{Lt}\int_0^t ds\,\bigg\lvert\sum_{j=1}^L(-1)^j\bra{\psi(s)}\hat{n}_j\ket{\psi(s)}\bigg\rvert,
\end{align}
\end{subequations}
respectively, where $\ket{\psi(t)}=e^{-i\hat{H}t}\ket{\psi_0}$. 

Figure~\ref{fig:nonlocal}(a) shows the dynamics of the gauge violation for a fixed gauge-breaking strength $\lambda$ at various values of the protection strength $V$, as calculated through ED. At early times, the gauge violation grows $\propto\lambda^2t^2$ as predicted by time-dependent perturbation theory \cite{Halimeh2020a}. After this initial growth, we see two distinct behaviors. At small $V$, the gauge violation is not suppressed, but rather grows to a maximal value at late times. However, at sufficiently large $V$, we see that the gauge violation plateaus at a timescale $\propto 1/V$ to a value $\propto\lambda^2/V^2$, in accordance with degenerate perturbation theory \cite{Halimeh2020a}, up to indefinite evolution times. Indeed, adapting results on slow heating in periodically driven systems \cite{abanin2017rigorous}, LPG protection with a \textit{rational} compliant sequence can be shown to stabilize gauge invariance up to times exponential in $V$, as we derive analytically in Appendix~\ref{sec:renormalized}. 

The long-time gauge violation as a function of $J/V$ is shown in Fig.~\ref{fig:nonlocal}(b). There, the two-regime behavior is clear in case of a compliant sequence. The long-time gauge violation goes from an uncontrolled-error regime at small $V$ to a controlled-error regime at sufficiently large $V$, at which it scales $\propto\lambda^2/V^2$. When it comes to the noncompliant sequence $c_j=[6(-1)^j+5]/11$, however, the violation does not enter a controlled-error regime, instead remaining above a minimum value no matter how large $V$ is. This is directly related to the nonlocal error term $\hat{H}_1^\text{nloc}$, which creates transitions between the few gauge-invariant sectors from which the LPG protection cannot isolate the target sector in the case of a noncompliant sequence. However, as we will show later, the noncompliant sequence is very powerful against local errors.

\begin{figure}[t!]
	\centering
	\includegraphics[width=.48\textwidth]{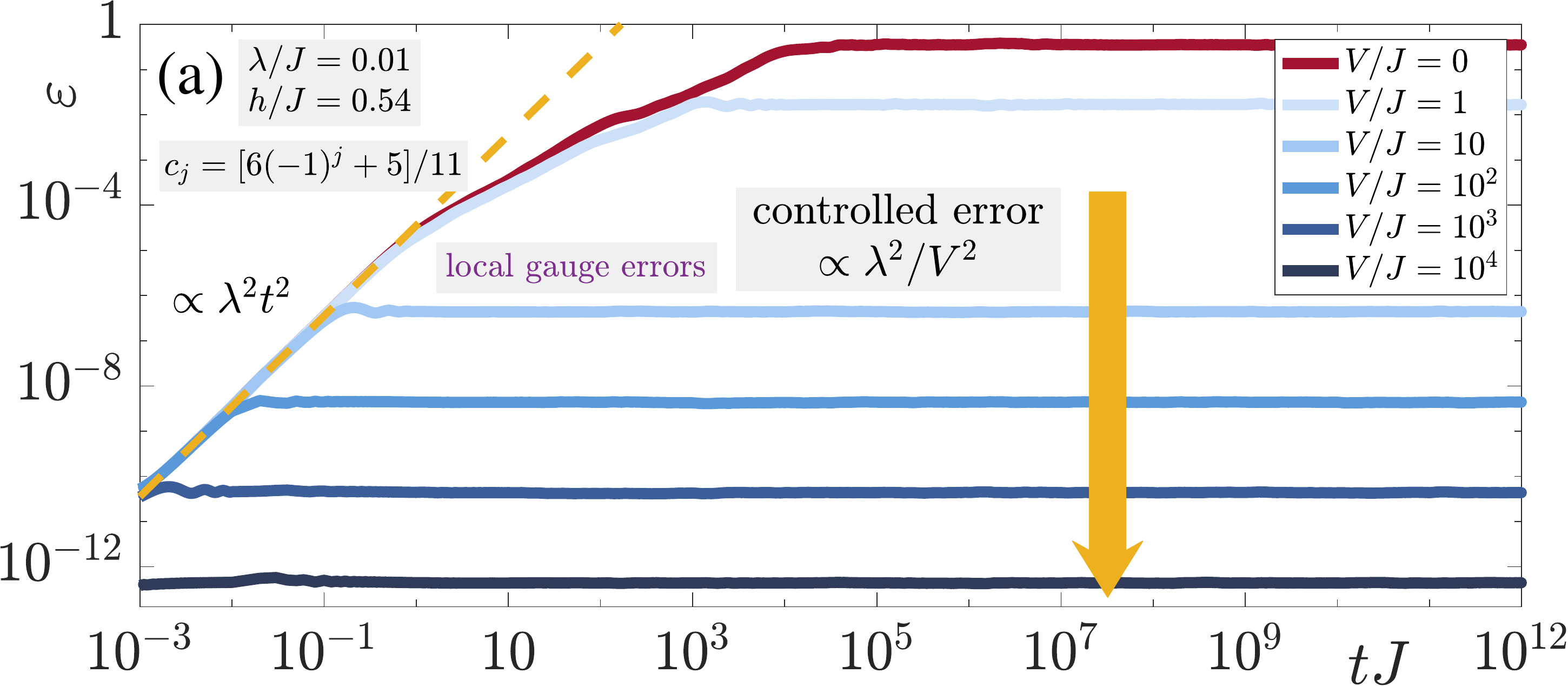}\\
	\vspace{1.1mm}
	\includegraphics[width=.48\textwidth]{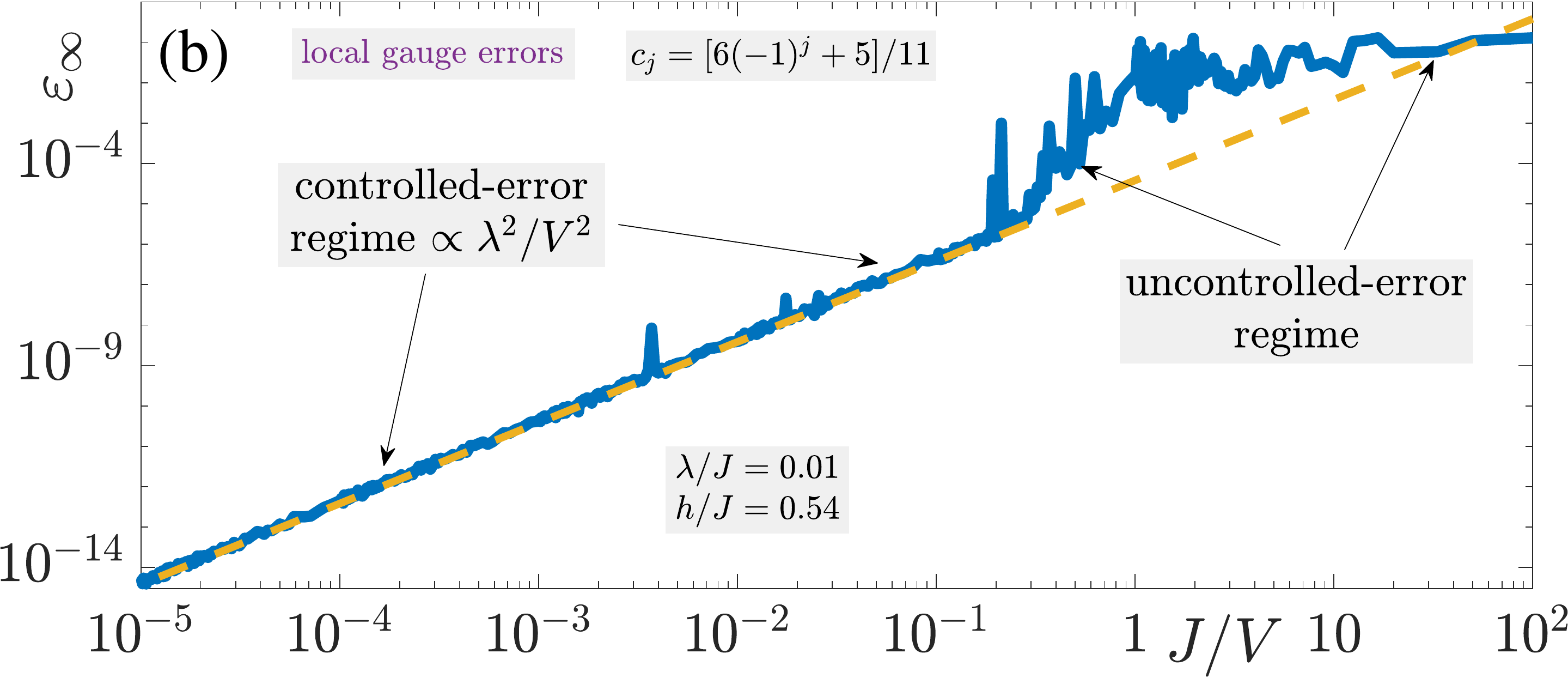}\\
	\vspace{1.1mm}
	\includegraphics[width=.48\textwidth]{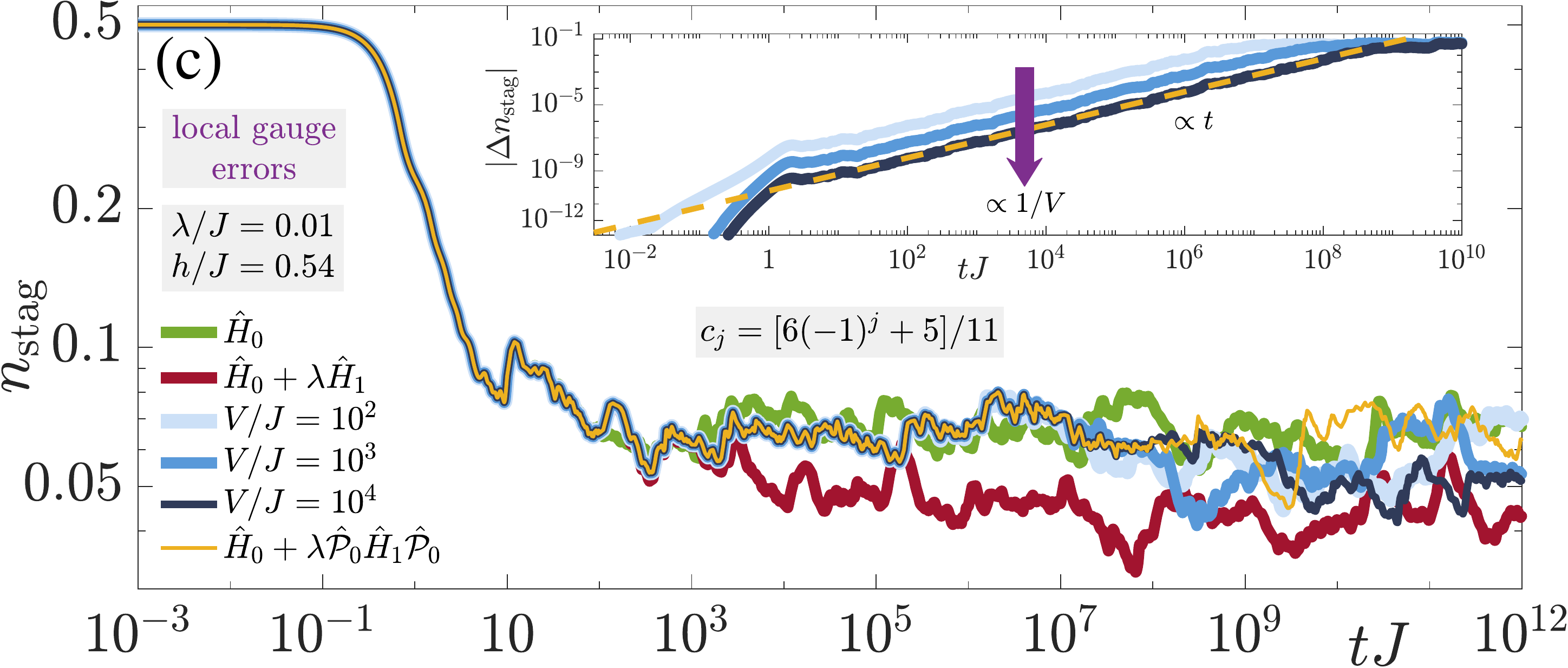}
	\caption{(Color online). Same as Fig.~\ref{fig:nonlocal} but with only local error terms given in Eq.~\eqref{eq:H1} and employing only the noncompliant sequence $c_j=[6(-1)^j+5]/11$ in the LPG protection of Eq.~\eqref{eq:LPGprotection}. The qualitative picture is identical to that of Fig.~\ref{fig:nonlocal} for the case of nonlocal errors and LPG protection with a compliant sequence, meaning that simplified sequences can reliably protect against experimentally relevant local error terms.}
	\label{fig:local} 
\end{figure}

As derived analytically in Appendix~\ref{sec:adjusted} through the quantum Zeno effect, we prove that the dynamics of local observables under the faulty theory $\hat{H}$ is faithfully reproduced by an adjusted gauge theory $\hat{H}_\text{adj}=\hat{H}_0+\lambda\hat{\mathcal{P}}_0(\hat{H}_1+\hat{H}_1^\text{nloc})\hat{\mathcal{P}}_0$, where $\hat{\mathcal{P}}_0$ is the projector onto the target sector. This occurs up to an error upper bound $\propto tV_0^2L^2/V$, yielding a timescale $\tau_\text{adj}\propto V/(V_0L)^2$, where $V_0$ is an energy constant depending on the microscopic parameters $\lambda/J$ and $h/J$. We find numerically that this is indeed the case for the staggered boson number under LPG protection with a compliant sequence as shown in Fig.~\ref{fig:nonlocal}(c). In the inset, the error in the dynamics under the faulty theory $\hat{H}$ with respect to $\hat{H}_\text{adj}$ grows linearly in time and is suppressed $\propto1/V$. It is to be noted here that although $\hat{H}_\text{adj}$ is generally different from the ideal gauge theory $\hat{H}_0$, it nevertheless has an exact local gauge symmetry.

\subsection{Experimentally relevant local gauge errors} We now demonstrate the efficacy of LPG protection with an experimentally feasible periodic noncompliant sequence $c_j$, in the case of the local gauge-breaking terms of Eq.~\eqref{eq:H1}. The faulty theory is now described by $\hat{H}=\hat{H}_0+\lambda \hat{H}_1+V\sum_j\hat{W}_j[6(-1)^j+5]/11$, and we quench again the staggered-matter initial state $\ket{\psi_0}$.

The dynamics of the gauge violation in Fig.~\ref{fig:local}(a) demonstrates reliable stabilization of gauge invariance with a plateau $\propto\lambda^2/V^2$ beginning at $t\propto 1/V$ and persisting over indefinite times at large enough $V$. Indeed, the transition from an uncontrolled to a controlled-error regime displayed in Fig.~\ref{fig:local}(b) occurs already at small values of $V\sim5J$, which is readily accessible in quantum-simulation setups \cite{Hauke2013,Schweizer2019,Yang2020}. The dynamics of $\hat{n}_\text{stag}$ in Fig.~\ref{fig:local}(c) is faithfully reproduced by the adjusted gauge theory $\hat{H}_0+\lambda\hat{\mathcal{P}}_0\hat{H}_1\hat{\mathcal{P}}_0$ up to the timescale $\tau_\text{adj}\propto V/(V_0L)^2$, with an error growing linearly in time and exhibiting a suppression $\propto 1/V$, as predicted analytically in Appendix~\ref{sec:adjusted}.

\begin{figure}[t!]
	\centering
	\includegraphics[width=.48\textwidth]{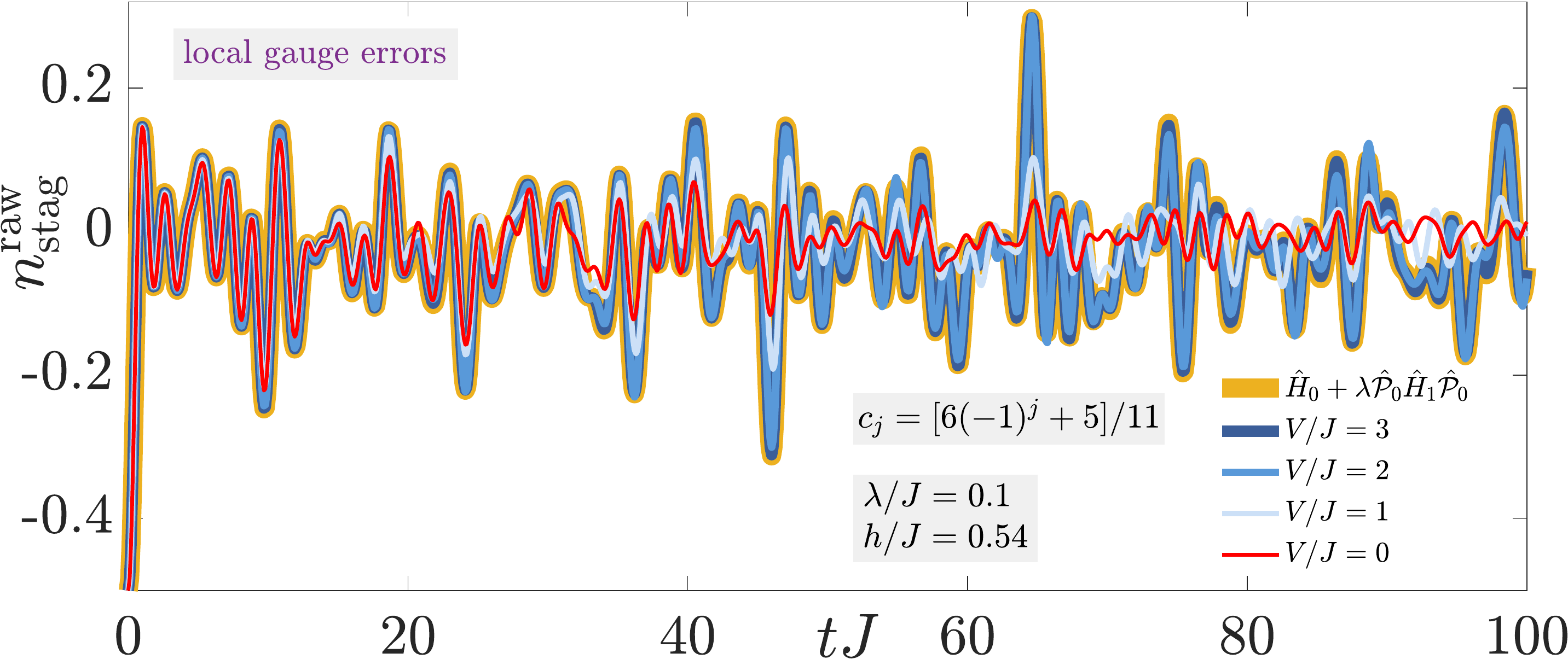}
	\caption{(Color online). Dynamics of the staggered boson number $\hat{n}_\text{stag}^\text{raw}(t)=\sum_j(-1)^j\bra{\psi(t)}\hat{n}_j\ket{\psi(t)}/L$ under the faulty gauge theory $\hat{H}=\hat{H}_0+\lambda \hat{H}_1+V\sum_j\hat{W}_j[6(-1)^j+5]/11$ demonstrates that LPG protection gives rise to an adjusted gauge theory $\hat{H}_0+\lambda\hat{\mathcal{P}}_0\hat{H}_1\hat{\mathcal{P}}_0$ during all experimentally relevant evolution times already at $V=2J$ and $\lambda=0.1J$, well within the accessible parameter range of state-of-the-art QSM devices.}
	\label{fig:experiment} 
\end{figure}

Within state-of-the-art quantum-simulation setups, it is possible to set $\lambda\sim0.1J$ and $V/\lambda\sim\mathcal{O}(3-28)$ \cite{Hauke2013,Schweizer2019,Yang2020}. Restricting our dynamics within experimentally feasible evolution times $t\lesssim100/J$, we find in Fig.~\ref{fig:experiment} that the staggered boson occupation is reliably reproduced by the adjusted gauge theory for $V/J=2$ with $\lambda/J=0.1$, i.e., well within the range of experimentally accessible parameters. This bodes well for ongoing efforts to stabilize local symmetries in quantum simulations of LGTs.

It is worth mentioning that in the $(1+1)-$D $\mathbb{Z}_2$ LGT, the LPG term given in Eq.~\eqref{eq:pseudoGen} is comprised of a single-body term, which is straightforward to realize in QSM setups, and of a two-body term, which can be reliably engineered using density-density interactions that, for e.g., naturally arise in ultracold-atom setups, where they are readily tuned using Feshbach resonances \cite{Bloch2008}, or in Rydberg arrays through dipole-dipole interactions \cite{Browaeys_review}.

\section{$(2+1)-$D $\mathbb{Z}_2$ lattice gauge theory}\label{sec:2D}
\begin{figure}[htp]
	\centering
	\includegraphics[width=.23\textwidth]{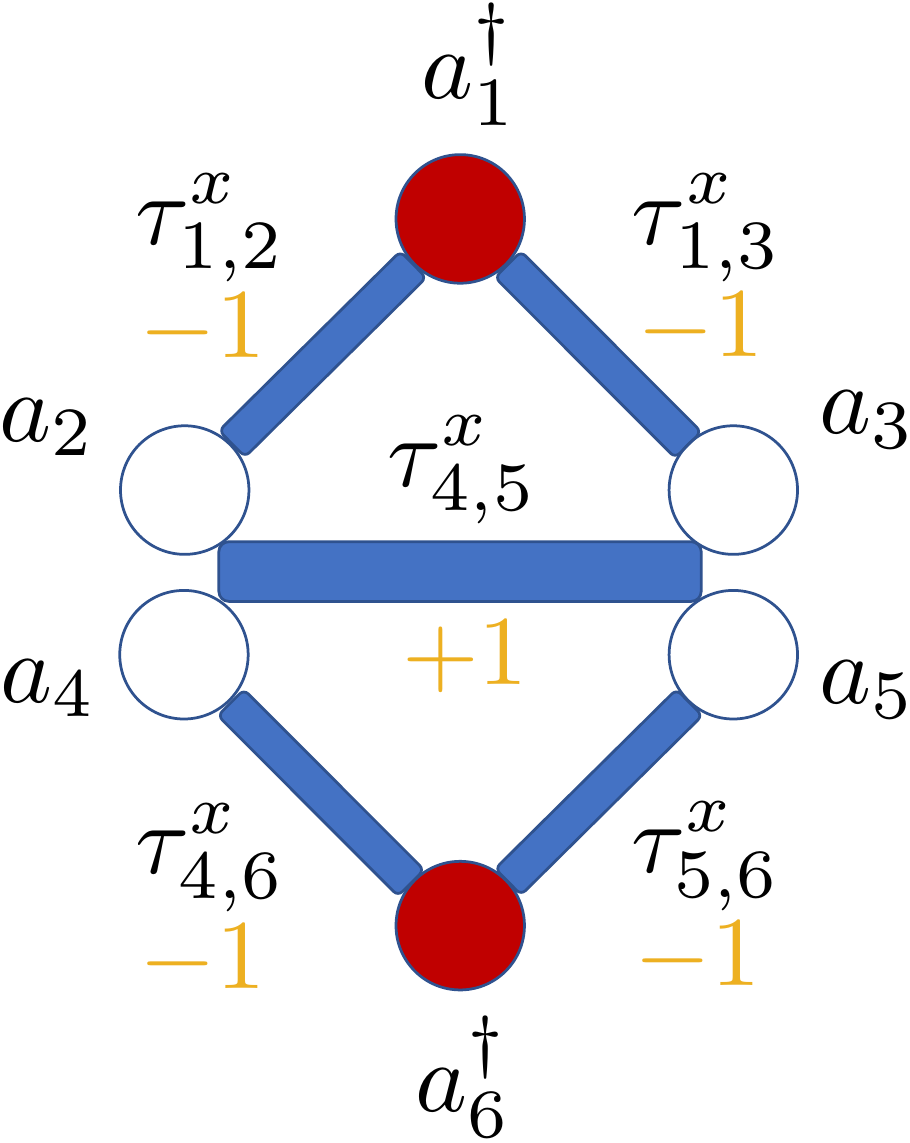}
	\caption{(Color online). $(2+1)-$D $\mathbb{Z}_2$ gauge theory on a triangular lattice with $L=6$ matter sites and $L_\ell=5$ gauge links. Circles indicate matter sites, with red circles denoting single occupation of hard-core bosons, while white circles are empty matter sites. The electric fields on the links between matter sites are initialized at one of their eigenvalues $\pm1$ (yellow). Note how the link between matter sites $2$ and $3$ is the same link as that between matter sites $4$ and $5$, i.e., $\hat{\tau}_{2,3}^{\{x,y,z\}}=\hat{\tau}_{4,5}^{\{x,y,z\}}$.}
	\label{fig:toric}
\end{figure}

We now show that the LPG protection scheme is not limited to strictly one-dimensional settings. To this end we consider a minimal $\mathbb{Z}_2$ LGT on a small triangular lattice shown in Fig.~\ref{fig:toric}, and described by the Hamiltonian \cite{homeier2020mathbbz2}
\begin{align}
\hat{H}_0=\sum_{\mathfrak{P},\langle l,j\rangle_\mathfrak{P}}\bigg(J\hat{a}_l^\dagger \hat{\tau}_{l,j}^z\hat{a}_j-\frac{h}{2}\hat{\tau}^x_{l,j}\bigg),
\end{align}
with the constraint that there is only a single link at the common edge of the plaquettes $\mathfrak{P}$, i.e., $\hat{\tau}^{\{x,y,z\}}_{2,3}=\hat{\tau}^{\{x,y,z\}}_{4,5}$. Gauge invariance is encoded by two types of generators. The first is $\hat{G}_j$ at a local constraint residing in only one plaquette and denoted by the matter site $j$ and its neighboring links, which is identical to its counterpart in $(1+1)-$D. The second is $\hat{G}_{l,j}$ at a local constraint shared by two plaquettes with eigenvalues $g_{l,j}=\pm1$, defined at a local constraint denoted by the matter site $l$ and its neighboring link on one plaquette and the matter site $j$ and its neighboring link on the second, along with the neighboring link common to both plaquettes. For clarity, we list them here explicitly:
\begin{subequations}
\begin{align}\label{eq:fullGen2D}
	&\hat{G}_1=(-1)^{\hat{n}_1}\hat{\tau}^x_{1,2}\hat{\tau}^x_{1,3},\\
	&\hat{G}_6=(-1)^{\hat{n}_6}\hat{\tau}^x_{4,6}\hat{\tau}^x_{5,6},\\
	&\hat{G}_{2,4}=(-1)^{\hat{n}_2+\hat{n}_4}\hat{\tau}^x_{1,2}\hat{\tau}^x_{4,5}\hat{\tau}^x_{4,6},\\
	&\hat{G}_{3,5}=(-1)^{\hat{n}_3+\hat{n}_5}\hat{\tau}^x_{1,3}\hat{\tau}^x_{4,5}\hat{\tau}^x_{5,6}.
\end{align}
\end{subequations}

To construct LPG terms with only up to two-body interactions for this system, we make a general ansatz for $\hat{W}_j$. This ansatz only contains couplings between $\hat{\tau}^x$ and $\hat{n}$ associated with a given vertex and treats all $\mathbb{Z}_2$ electric field terms on equal footing. Allowing for arbitrary interaction strengths and requiring the eigenenergies of the constructed interaction term to collapse in a given target gauge sector yields possible solutions for the form of $\hat{W}_j$. 

The LPGs we find for the full generators in Eq.~\eqref{eq:fullGen2D} are
\begin{subequations}
\begin{align}\label{eq:pseudoGen2D}
	&\hat{W}_1(g_1^\text{tar})=\hat{\tau}^x_{1,2}\hat{\tau}^x_{1,3}+2g_1^\text{tar}\hat{n}_1,\\
	&\hat{W}_6(g_6^\text{tar})=\hat{\tau}^x_{4,6}\hat{\tau}^x_{5,6}+2g_6^\text{tar}\hat{n}_6,\\
	&\hat{W}_{2,4}(g_{2,4}^\text{tar})=\hat{\tau}^x_{1,2}\hat{\tau}^x_{4,5}\hat{\tau}^x_{4,6}+2g_{2,4}^\text{tar}(\hat{n}_2+\hat{n}_4-2\hat{n}_2\hat{n}_4),\\
	&\hat{W}_{3,5}(g_{3,5}^\text{tar})=\hat{\tau}^x_{1,3}\hat{\tau}^x_{4,5}\hat{\tau}^x_{5,6}+2g_{3,5}^\text{tar}(\hat{n}_3+\hat{n}_5-2\hat{n}_3\hat{n}_5),
\end{align}
\end{subequations}
respectively, which act identically to their full counterparts in the target sector, as shown in Tables~\ref{Table1D} and~\ref{Table2D}. Indeed, whenever the eigenvalue $w_{l,j}(g_{l,j}^\text{tar})$ of $\hat{W}_{l,j}(g_{l,j}^\text{tar})$ equals $g_{l,j}^\text{tar}$, then so does the eigenvalue $g_{l,j}$ of $\hat{G}_{l,j}$, and vice versa: $w_{l,j}(g_{l,j}^\text{tar})=g_{l,j}^\text{tar}\iff g_{l,j}=g_{l,j}^\text{tar}$.

\begin{table}[h!]
	\centering
	\begin{tabular}{|| c | c | c | c | c || c | c | c ||} 
		\hline
		$\hat{n}_2$ & $\hat{n}_3$ & $\hat{\tau}_{1,2}^x$ & $\hat{\tau}_{4,5}^x$ & $\hat{\tau}_{4,6}^x$ & $\hat{G}_{2,4}$ & $\hat{W}_{2,4}(g_{2,4}^\text{tar}={\color{myred}-1})$ & $\hat{W}_{2,4}(g_{2,4}^\text{tar}={\color{mygreen}+1})$ \\ [0.5ex] 
		\hline\hline
		$0$ & $0$ &  $-1$ & $-1$ & $-1$ & \color{myred}$-1$ & \color{myred}$-1$ & $-1$\\ 
		\hline
		$0$ & $0$ &  $-1$ & $-1$ & $+1$ & \color{mygreen}$+1$ & $+1$ & \color{mygreen}$+1$\\ 
		\hline
		$0$ & $0$ &  $-1$ & $+1$ & $-1$ & \color{mygreen}$+1$ & $+1$ & \color{mygreen}$+1$\\ 
		\hline
		$0$ & $0$ &  $-1$ & $+1$ & $+1$ & \color{myred}$-1$ & \color{myred}$-1$ & $-1$\\ 
		\hline
		$0$ & $0$ &  $+1$ & $-1$ & $-1$ & \color{mygreen}$+1$ & $+1$ & \color{mygreen}$+1$\\ 
		\hline
		$0$ & $0$ &  $+1$ & $-1$ & $+1$ & \color{myred}$-1$ & \color{myred}$-1$ & $-1$\\ 
		\hline
		$0$ & $0$ &  $+1$ & $+1$ & $-1$ & \color{myred}$-1$ & \color{myred}$-1$ & $-1$\\ 
		\hline
		$0$ & $0$ &  $+1$ & $+1$ & $+1$ & \color{mygreen}$+1$ & $+1$ & \color{mygreen}$+1$\\ 
		\hline
		$0$ & $1$ &  $-1$ & $-1$ & $-1$ & \color{mygreen}$+1$ & $-3$ & \color{mygreen}$+1$\\ 
		\hline
		$0$ & $1$ &  $-1$ & $-1$ & $+1$ & \color{myred}$-1$ & \color{myred}$-1$ & $+3$\\ 
		\hline
		$0$ & $1$ &  $-1$ & $+1$ & $-1$ & \color{myred}$-1$ & \color{myred}$-1$ & $+3$\\ 
		\hline
		$0$ & $1$ &  $-1$ & $+1$ & $+1$ & \color{mygreen}$+1$ & $-3$ & \color{mygreen}$+1$\\ 
		\hline
		$0$ & $1$ &  $+1$ & $-1$ & $-1$ & \color{myred}$-1$ & \color{myred}$-1$ & $+3$\\ 
		\hline
		$0$ & $1$ &  $+1$ & $-1$ & $+1$ & \color{mygreen}$+1$ & $-3$ & \color{mygreen}$+1$\\ 
		\hline
		$0$ & $1$ &  $+1$ & $+1$ & $-1$ & \color{mygreen}$+1$ & $-3$ & \color{mygreen}$+1$\\ 
		\hline
		$0$ & $1$ &  $+1$ & $+1$ & $+1$ & \color{myred}$-1$ & \color{myred}$-1$ & $+3$\\ 
		\hline
		$1$ & $0$ &  $-1$ & $-1$ & $-1$ & \color{mygreen}$+1$ & $-3$ & \color{mygreen}$+1$\\ 
		\hline
		$1$ & $0$ &  $-1$ & $-1$ & $+1$ & \color{myred}$-1$ & \color{myred}$-1$ & $+3$\\ 
		\hline
		$1$ & $0$ &  $-1$ & $+1$ & $-1$ & \color{myred}$-1$ & \color{myred}$-1$ & $+3$\\ 
		\hline
		$1$ & $0$ &  $-1$ & $+1$ & $+1$ & \color{mygreen}$+1$ & $-3$ & \color{mygreen}$+1$\\ 
		\hline
		$1$ & $0$ &  $+1$ & $-1$ & $-1$ & \color{myred}$-1$ & \color{myred}$-1$ & $+3$\\ 
		\hline
		$1$ & $0$ &  $+1$ & $-1$ & $+1$ & \color{mygreen}$+1$ & $-3$ & \color{mygreen}$+1$\\ 
		\hline
		$1$ & $0$ &  $+1$ & $+1$ & $-1$ & \color{mygreen}$+1$ & $-3$ & \color{mygreen}$+1$\\ 
		\hline
		$1$ & $0$ &  $+1$ & $+1$ & $+1$ & \color{myred}$-1$ & \color{myred}$-1$ & $+3$\\ 
		\hline
		$1$ & $1$ &  $-1$ & $-1$ & $-1$ & \color{myred}$-1$ & \color{myred}$-1$ & $-1$\\ 
		\hline
		$1$ & $1$ &  $-1$ & $-1$ & $+1$ & \color{mygreen}$+1$ & $+1$ & \color{mygreen}$+1$\\ 
		\hline
		$1$ & $1$ &  $-1$ & $+1$ & $-1$ & \color{mygreen}$+1$ & $+1$ & \color{mygreen}$+1$\\ 
		\hline
		$1$ & $1$ &  $-1$ & $+1$ & $+1$ & \color{myred}$-1$ & \color{myred}$-1$ & $-1$\\ 
		\hline
		$1$ & $1$ &  $+1$ & $-1$ & $-1$ & \color{mygreen}$+1$ & $+1$ & \color{mygreen}$+1$\\ 
		\hline
		$1$ & $1$ &  $+1$ & $-1$ & $+1$ & \color{myred}$-1$ & \color{myred}$-1$ & $-1$\\ 
		\hline
		$1$ & $1$ &  $+1$ & $+1$ & $-1$ & \color{myred}$-1$ & \color{myred}$-1$ & $-1$\\ 
		\hline
		$1$ & $1$ &  $+1$ & $+1$ & $+1$ & \color{mygreen}$+1$ & $+1$ & \color{mygreen}$+1$\\ 
		[1ex] 
		\hline
	\end{tabular}
	\caption{Eigenvalues $w_{2,4}(g_{2,4}^\text{tar})$ and $g_{2,4}$ of the inter-plaquette local pseudogenerator $\hat{W}_{2,4}(g_{2,4}^\text{tar})$ and the corresponding full generator $\hat{G}_{2,4}$, respectively, are identical in the target sector $g_{2,4}^\text{tar}$, such that $w_{2,4}(g_{2,4}^\text{tar})=g_{2,4}^\text{tar}\iff g_{2,4}=g_{2,4}^\text{tar}$.}
	\label{Table2D}
\end{table}

Experimentally relevant local gauge-breaking errors for this model have been determined to be of the form
\begin{align}\nonumber
\lambda \hat{H}_1=\sum_{\mathfrak{P},\langle l,j\rangle_\mathfrak{P}}\Big[&\beta_1\hat{a}_l^\dagger \hat{a}_j+\beta_2\hat{\tau}^z_{l,j}+\beta_3\big(\hat{n}_l+\hat{n}_j\big)\hat{\tau}^z_{l,j}\\\label{eq:loc2D}
&+\beta_4\hat{n}_l\hat{n}_j\hat{\tau}^z_{l,j}\Big],
\end{align}
with $\beta_1=0.06$ and $\beta_2=\beta_3=\beta_4=0.01$ \cite{homeier2020mathbbz2}, although we have checked that our qualitative picture remains the same for other values of $\beta_{1\ldots4}$. Furthermore, in order to further scrutinize the LPG protection in $(2+1)-$D, we have also included the experimentally very unlikely nonlocal error
\begin{align}\label{eq:nloc2D}
\lambda \hat{H}_1^\text{nloc}=\lambda\sum_{\xi=\pm1}\prod_{\mathfrak{P},\langle l,j\rangle_\mathfrak{P}}\big(\mathds{1}+\xi\hat{\tau}^z_{l,j}\big).
\end{align}

The LPG protection term used to suppress gauge violations due to these errors is described by
\begin{align}\nonumber
V\hat{H}_W=&V\Big\{c_1\big[\hat{W}_1(g_1^\text{tar})-g_1^\text{tar}\big]+c_2\big[\hat{W}_2(g_2^\text{tar})-g_2^\text{tar}\big]\\\label{eq:LPG2D}
&+c_3\big[\hat{W}_{2,4}(g_{2,4}^\text{tar})-g_{2,4}^\text{tar}\big]+c_4\big[\hat{W}_{3,5}(g_{3,5}^\text{tar})-g_{3,5}^\text{tar}\big]\Big\},
\end{align}
with the noncompliant sequence $c_j\in\{-1,2,-3,5\}/5$. As we will see, for this $2$D geometry, even a noncompliant sequence renders LPG protection powerful enough to suppress such extreme nonlocal gauge-breaking errors.

\begin{figure}[htp]
	\centering
	\includegraphics[width=.48\textwidth]{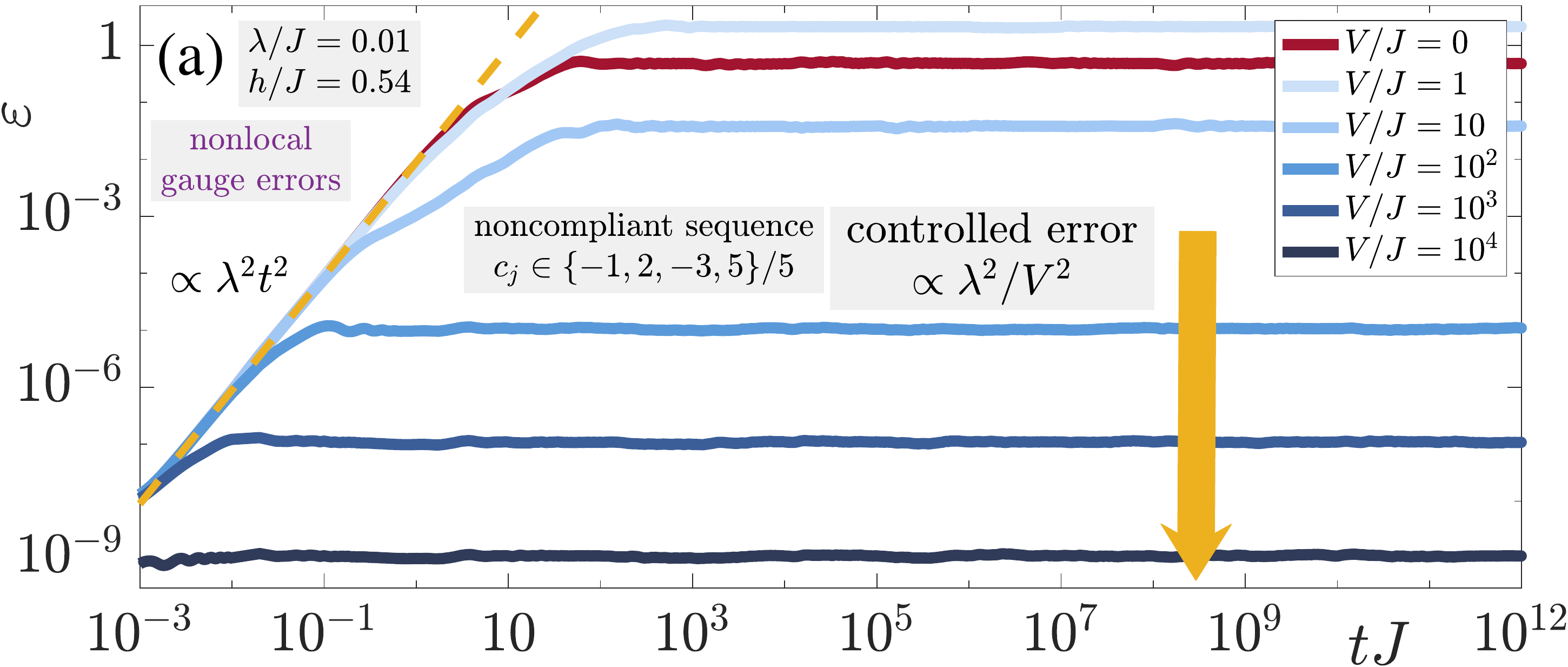}\\
	\vspace{1.1mm}
	\includegraphics[width=.48\textwidth]{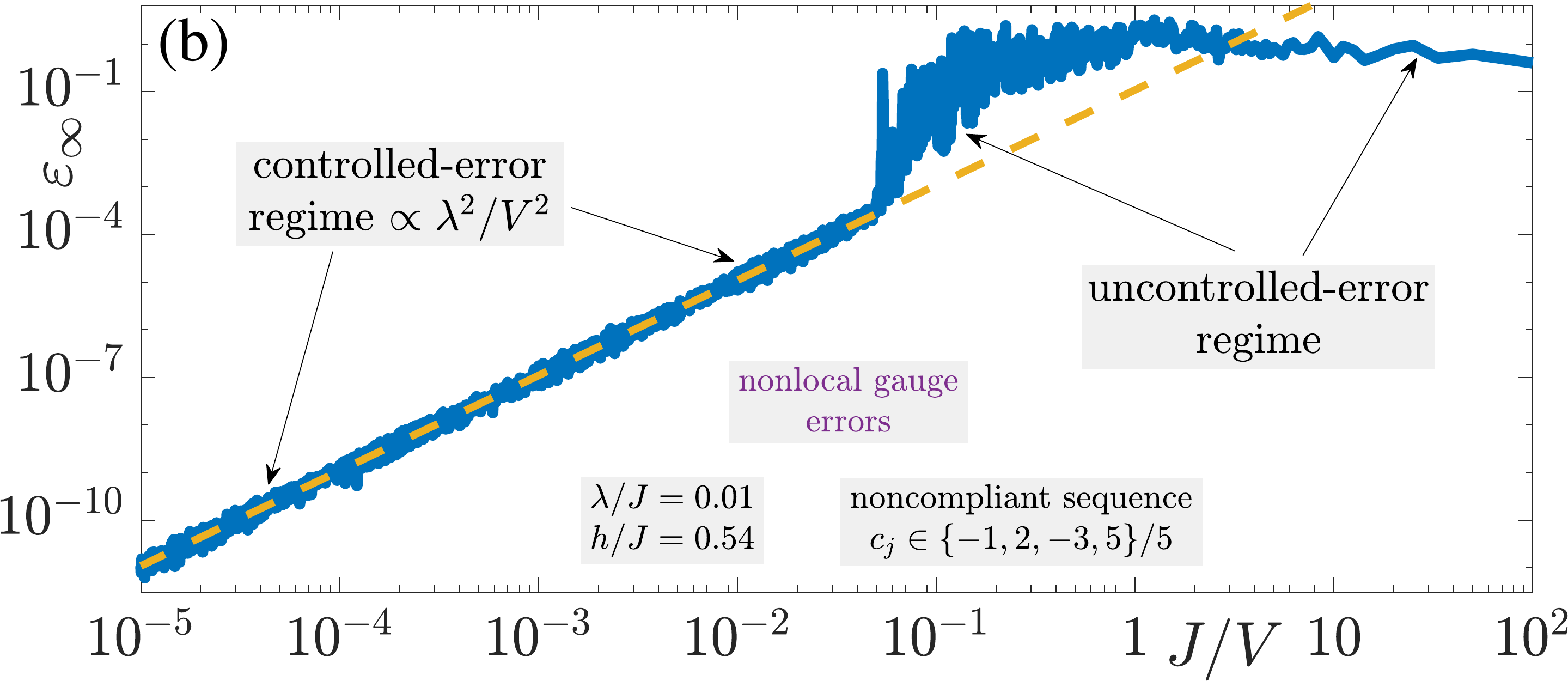}\\
	\vspace{1.1mm}
	\includegraphics[width=.48\textwidth]{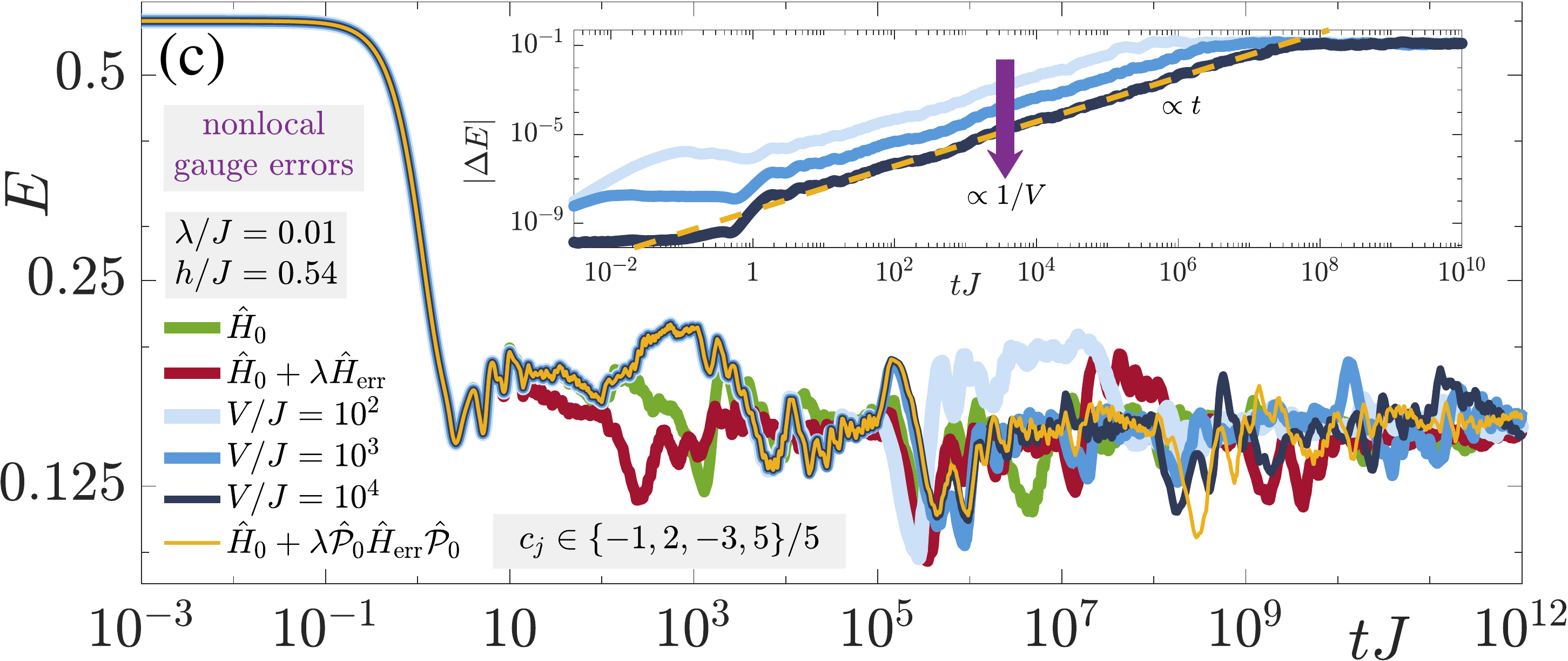}
	\caption{(Color online). $(2+1)-$D $\mathbb{Z}_2$ LGT on a triangular lattice with gauge-breaking terms $\hat{H}_\text{err}=\hat{H}_1+\hat{H}_1^\text{nloc}$ given in Eqs.~\eqref{eq:loc2D} and~\eqref{eq:nloc2D}. LPG protection with a noncompliant sequence, see Eq.~\eqref{eq:LPG2D}, is used to stabilize gauge invariance. (a) Gauge-violation dynamics at sufficiently large $V$ settles into a plateau $\propto\lambda^2/V^2$ that begins at a timescale $\propto1/V$ and lasts up to all accessible times in ED. It is remarkable that this occurs despite the LPG-protection sequence being noncompliant, which seems unable to protect against such extremely nonlocal errors in $(1+1)-$D, see Fig.~\ref{fig:nonlocal}(b). (b) A two-regime picture emerges, with an uncontrolled long-time violation at small enough values of $V$, while at sufficiently large values of $V$ the long-time violation enters a regime of controlled error $\propto\lambda^2/V^2$. (c) LPG protection gives rise to the adjusted gauge theory $\hat{H}_\text{adj}=\hat{H}_0+\lambda\hat{\mathcal{P}}_0\hat{H}_\text{err}\hat{\mathcal{P}}_0$, which faithfully reproduces the dynamics of the electric field under the faulty theory up to a timescale $\tau_\text{adj}\propto V/(V_0L)^2$. As predicted analytically, the corresponding error grows linear in time and is suppressed as $\propto 1/V$.}
	\label{fig:2D} 
\end{figure}

We prepare our initial state $\ket{\psi_0}$ in the target sector $g_1^\text{tar}=g_6^\text{tar}=-1$ and $g_{2,4}^\text{tar}=g_{3,5}^\text{tar}=+1$ (see Fig.~\ref{fig:toric}), and quench with the faulty gauge theory $\hat{H}=\hat{H}_0+\lambda(\hat{H}_1+\hat{H}_1^\text{nloc})+V\hat{H}_W$ for $\lambda/J=0.01$ and $h/J=0.54$, although we have checked that our qualitative conclusions hold for other values of these parameters. We show the dynamics of the temporally averaged gauge violation, Eq.~\eqref{eq:viol}, for several values of $V$ (see legend) in Fig.~\ref{fig:2D}(a). Remarkably, we see at sufficiently large $V$ a suppression of the gauge violation, which enters a plateau at the timescale $\propto1/V$ with a value $\propto\lambda^2/V^2$ even when the sequence is noncompliant and the gauge-breaking error includes strongly nonlocal terms. We have not been able to find a noncompliant sequence that achieves this for the $(1+1)-$D model; we speculate that in higher dimensions the higher connectivity may further restrict how gauge violations spread \cite{Halimeh2020b,Halimeh2020c}. A scan of the long-time gauge violation as a function of $J/V$ also shows two distinct regimes. For small enough $V$, the violation cannot be directly related to the value of $V$, and falls into an uncontrolled-error regime. At sufficiently large $V$, we find that the gauge violation enters a controlled-error regime and behaves $\propto\lambda^2/V^2$.

Finally, we look in Fig.~\ref{fig:2D}(c) at the temporally averaged absolute electric field
\begin{align}
E=\frac{1}{2L_\ell t}\int_0^t ds\,\Big\lvert\sum_{\mathfrak{P},\langle l,j\rangle_\mathfrak{P}}\bra{\psi(s)}\hat{\tau}^x_{l,j}\ket{\psi(s)}\Big\rvert,
\end{align}
where $L_\ell=5$ is the number of links on the triangular lattice of Fig.~\ref{fig:toric}. The qualitative picture is the same as for our other results, with LPG protection giving rise to an adjusted gauge theory $\hat{H}_\text{adj}=\hat{H}_0+\lambda\hat{\mathcal{P}}_0(\hat{H}_1+\hat{H}_1^\text{nloc})\hat{\mathcal{P}}_0$ that faithfully reproduces the dynamics of $E$ under the faulty gauge theory within an error upper bound $\propto tV_0^2L^2/V$, i.e., up to a timescale $\tau_\text{adj}=V/(V_0L)^2$. The inset shows how the deviation of the dynamics under the faulty theory relative to that under the adjusted gauge theory scales $\propto1/V$ and grows linearly in time, which is within our analytic predictions (see Sec.~\ref{sec:adjusted}).

\section{Summary and outlook}\label{sec:summary}
We have introduced the concept of simplified local pseudogenerators (LPGs) that behave within the target sector identically to the actual generators of the gauge symmetry. This greatly simplifies experimental requirements compared to the implementation of the full generator to stabilize gauge invariance, as by construction the pseudogenerator has fewer particles per term than its full counterpart. We have demonstrated the efficacy of LPG protection in one and two spatial dimensions even under the severe case of nonlocal errors with support over the entire lattice, where it stabilized gauge invariance up to all accessible times in ED. We have also provided analytic predictions supporting these findings, and predicting the emergence of an adjusted gauge theory up to timescales polynomial in the LPG protection strength. Furthermore, we have shown that LPG protection provides robust stability of gauge invariance within experimentally accessible parameter regimes in current quantum simulators, which means LPGs should be a viable tool that can already be employed in such devices.

Even though we have focused in the main results on the $\mathbb{Z}_2$ LGT, which has a discrete spectrum, we emphasize that LPG protection is general and can be employed for other Abelian gauge theories in any dimension. An immediate future direction arising from our work is extending LPG protection to non-Abelian LGTs, where the concept of linear protection does not work in general \cite{Halimeh2021gauge} specifically because the local generators do not commute. It would be interesting to investigate whether commuting LPGs can be contrived that act within the target sector as the actual generators of the non-Abelian gauge symmetry.

\begin{acknowledgments}
We are grateful to Haifeng Lang for stimulating discussions, and for a meticulous reading of and valuable comments on our manuscript. This work is part of and supported by Provincia Autonoma di Trento, the ERC Starting Grant StrEnQTh (project ID 804305), the Google Research Scholar Award ProGauge, and Q@TN — Quantum Science and Technology in Trento. This research was supported in part by the National Science Foundation under Grant No.~NSF PHY-1748958. M.A.~and F.G.~acknowledge funding from the Deutsche Forschungsgemeinschaft (DFG, German Research Foundation) via Research Unit FOR 2414 under project number 277974659, and under Germany’s Excellence Strategy – EXC-2111 – 390814868. M.A.~further acknowledges funding from the European Research Council (ERC) under the European Union’s Horizon 2020 research and innovation program (grant agreement No.~803047). C.S.~has received funding from the European Union’s Framework Programme for Research and Innovation Horizon 2020 (2014-2020) under the Marie Sk\l odowska-Curie Grant Agreement No.\ 754388 (LMUResearchFellows) and from LMUexcellent, funded by the Federal Ministry of Education and Research (BMBF) and the Free State of Bavaria under the Excellence Strategy of the German Federal Government and the L\"ander.
\end{acknowledgments}

\appendix
\section{Supporting results for the $(1+1)-$D $\mathbb{Z}_2$ lattice gauge theory}\label{sec:supporting}
In this Appendix, we provide numerical results supporting the conclusions of the main text for the $(1+1)-$D $\mathbb{Z}_2$ LGT, by showcasing the efficacy of LPG protection compared to ``full'' energy-penalty protection, and by demonstrating its robustness to various initial conditions, model parameters, nonperturbative errors, and also to nonlocal gauge-breaking terms that simultaneously violate the global $\mathrm{U}(1)$ symmetry of boson-number conservation. 

\subsection{Comparison with full gauge protection}\label{sec:comp}

\begin{figure}[htp]
	\centering
	\includegraphics[width=.48\textwidth]{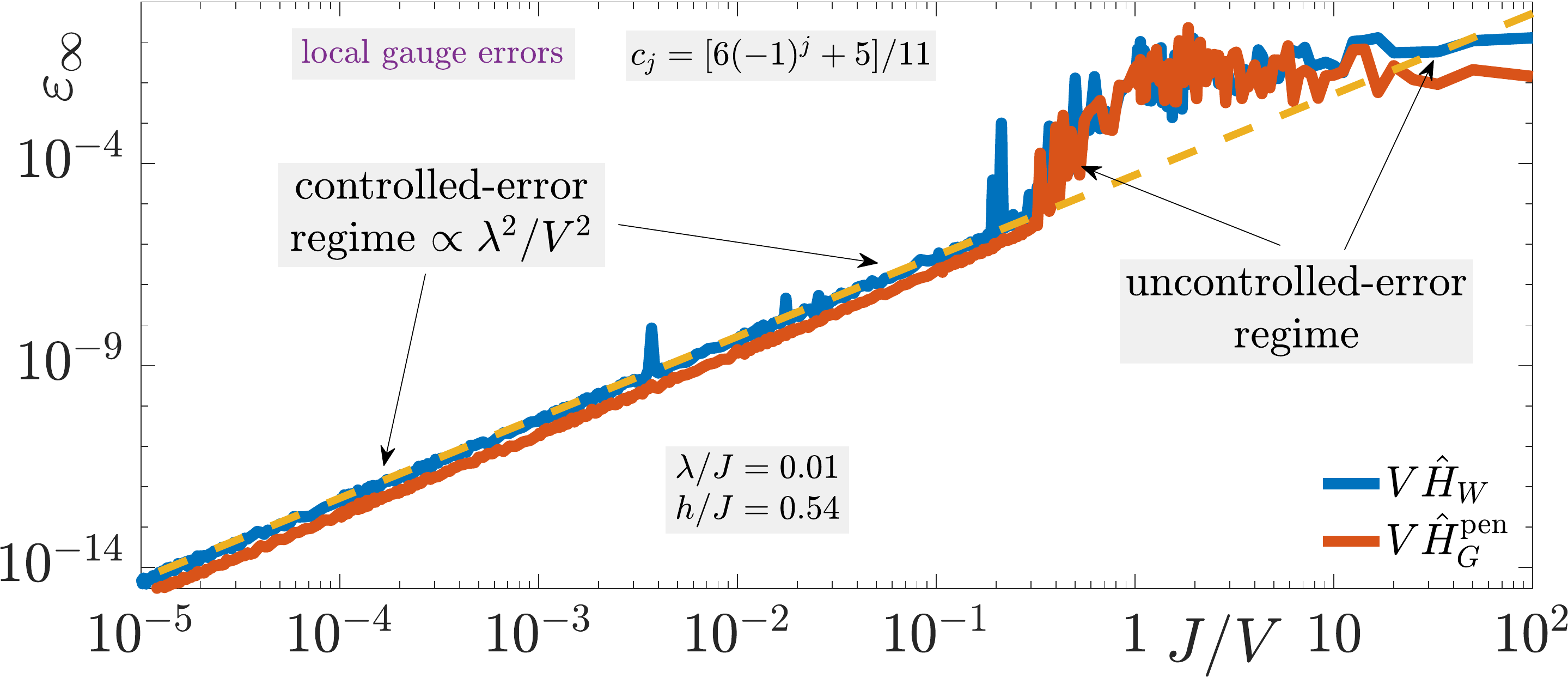}
	\caption{(Color online). Comparing LPG protection to full protection in the case of experimentally relevant local errors. The LPG protection sequence here is noncompliant, with $c_j=[6(-1)^j+5]/11$, but as shown in the main text, this is sufficient to protect against local gauge errors. Here we have chosen $\lambda/J=0.01$ and $h/J=0.54$, but we have checked that our results hold for other values of these parameters. Even though the full protection shows slightly better protection quantitatively, the LPG protection exhibits similar qualitative performance, with the transition from an uncontrolled-error to a controlled-error $\propto\lambda^2/V^2$ regime occurring at $V\gtrsim 5J$ compared to $V\gtrsim 3J$ for the full protection. In the case of LPG protection, certain resonances between the target sector and other gauge-invariant sectors are not fully controlled at certain values of $V$ within the controlled-error regime (see small ``spike'' at $J/V\approx3.7\times10^{-3}$), but nevertheless they are still reliably suppressed.}
	\label{fig:comp}
\end{figure}

It is interesting to compare the performance of the LPG protection of Eq.~\eqref{eq:LPGprotection} with a noncompliant sequence to that of the full protection $V\hat{H}_G^\text{pen}=V\sum_j(1-\hat{G}_j)$, where here the target sector is chosen to be $g_j^\text{tar}=+1$ as in the main text. For this purpose, we scan the ``infinite''-time gauge violation $\varepsilon_\infty=1-\lim_{t\to\infty}\sum_{j=1}^L\bra{\psi(t)}\hat{G}_j\ket{\psi(t)}/L$ as a function of $J/V$ under LPG protection with a noncompliant sequence and under full protection, in the presence of the experimentally relevant local errors given in Eq.~\eqref{eq:H1}. In our ED calculations, ``infinite'' time is chosen to be numerically anywhere between $t/J=10^5-10^{12}$, as the result is qualitatively independent of the value of $t\gtrsim10^5/J$. The results are shown in Fig.~\ref{fig:comp}, where the LPG protection exhibits qualitatively similar efficacy to the full protection. Indeed, in both cases we see a clear transition from an uncontrolled-error to a controlled-error regime where the steady-state value of the gauge violation scales $\propto\lambda^2/V^2$. We have chosen here the experimentally feasible noncompliant sequence $c_j=[6(-1)^j+5]/11$ for the LPG protection. Unlike the case of a compliant sequence, this does not isolate the target sector from \textit{all} other gauge-invariant sectors. This leads to imperfections at a few values of $V$ in the behavior of the infinite-time violation within the controlled-error regime, albeit the suppression of the violation is still remarkably reliable at these values as well. As such, this is quite encouraging news for ongoing experiments that the LPG protection with the experimentally feasible noncompliant periodic sequence can perform qualitatively as well as the full protection.

\subsection{Results for different initial states, model parameters, and error strengths}
\begin{figure}[htp]
	\centering
	\includegraphics[width=.48\textwidth]{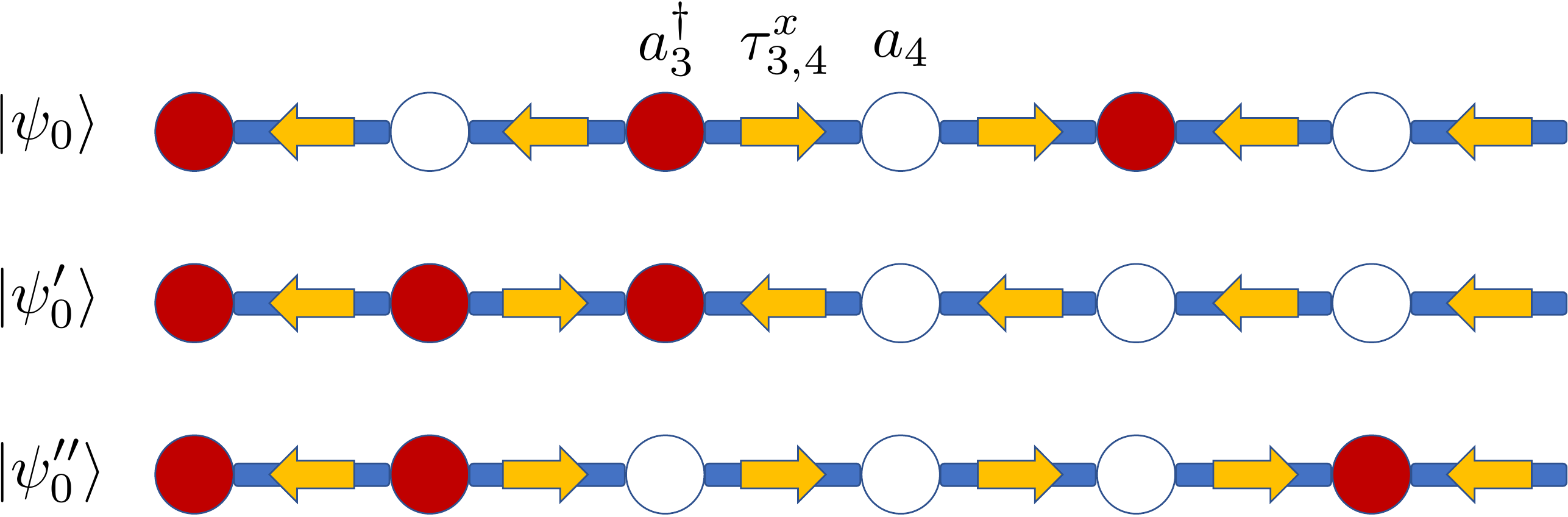}
	\caption{(Color online). Initial states used in our ED calculations. Circles represent matter sites, where red circles denote single hard-core boson occupation and white circles are empty matter sites. The yellow arrows on the links between matter sites denote the eigenvalue of the electric field $\hat{\tau}_{j,j+1}^x$ as $\pm1$ when pointing right (left). All three initial states are in the target sector $g_j^\text{tar}=+1,\,\forall j$. In the main text, we have focused on $\ket{\psi_0}$, although LPG protection offers reliable stabilization of gauge invariance independently of the initial state, as shown in Fig.~\ref{fig:diff}(a).}
	\label{fig:InitialStates}
\end{figure}

\begin{figure}[t!]
	\centering
	\includegraphics[width=.48\textwidth]{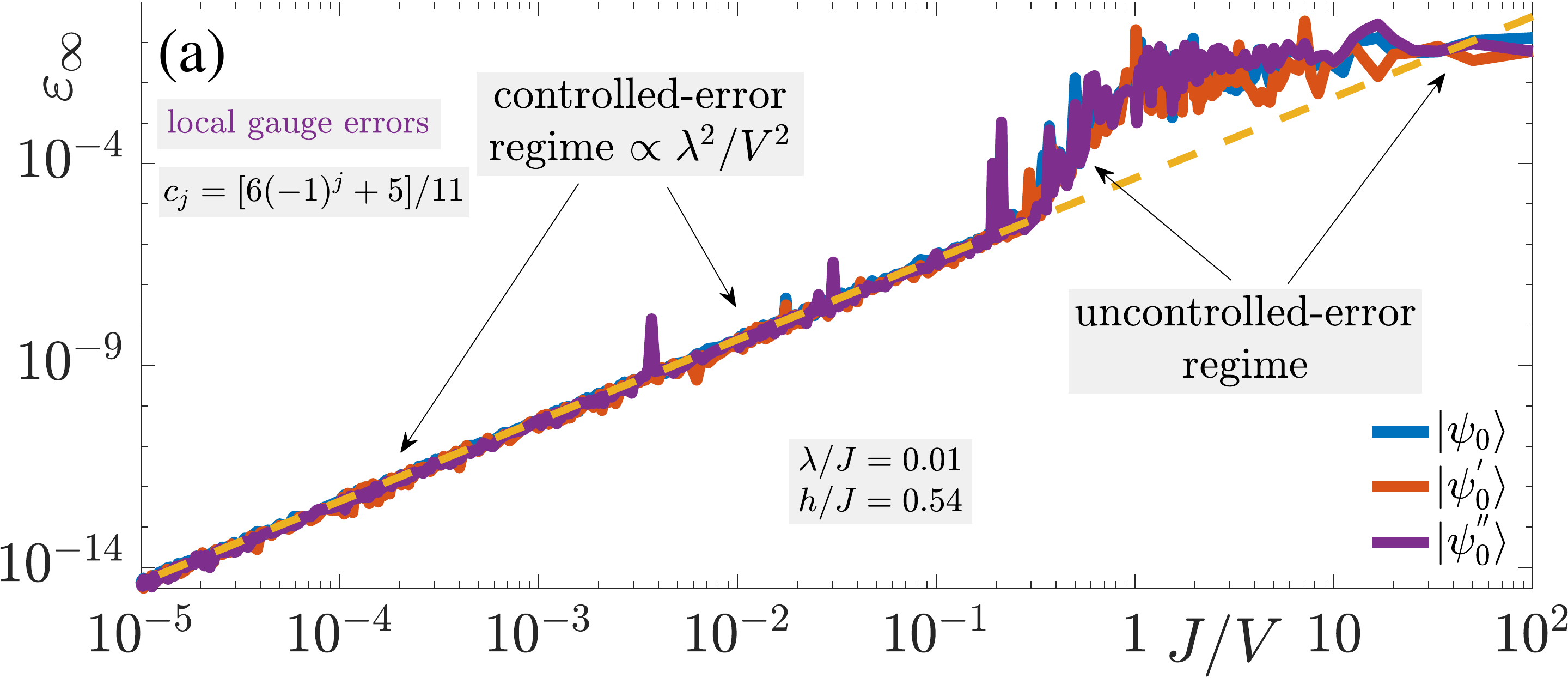}\\
	\vspace{1.1mm}
	\includegraphics[width=.48\textwidth]{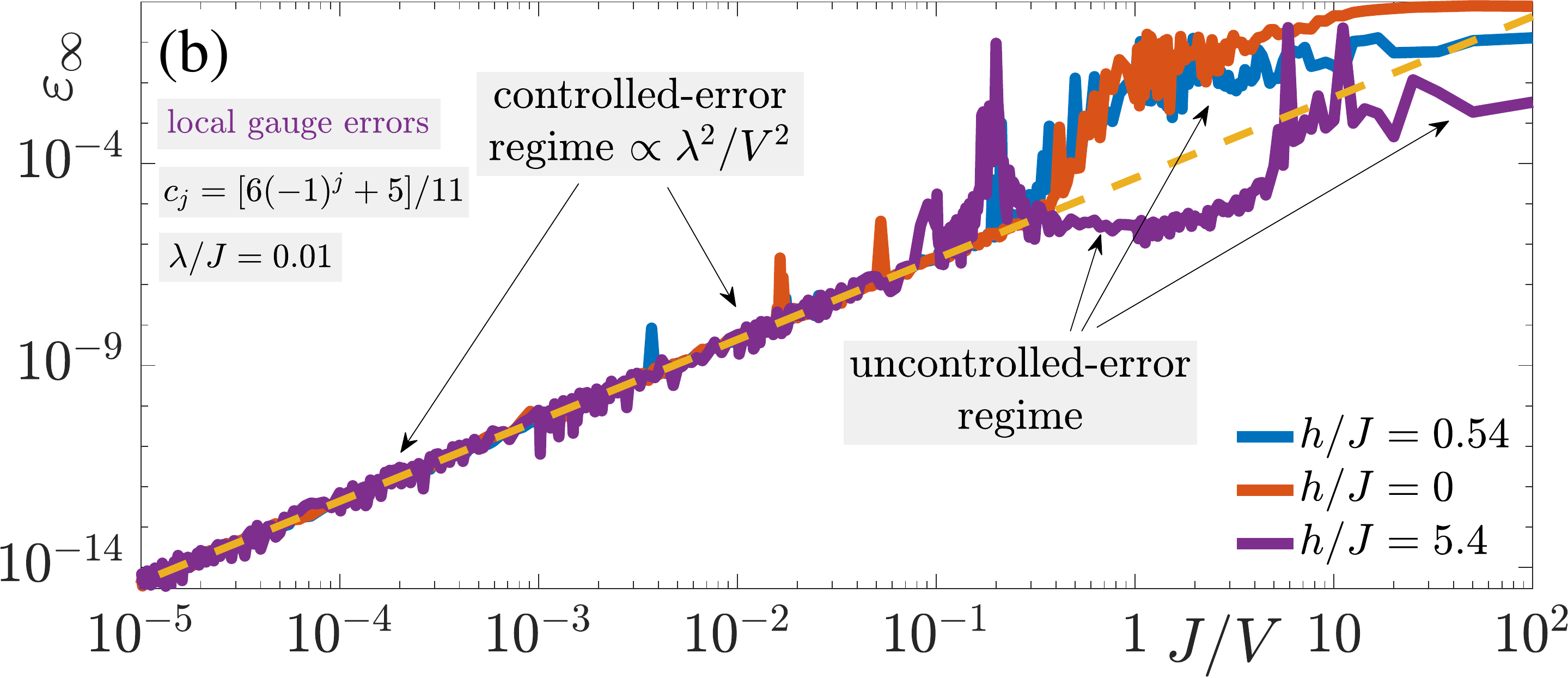}\\
	\vspace{1.1mm}
	\includegraphics[width=.48\textwidth]{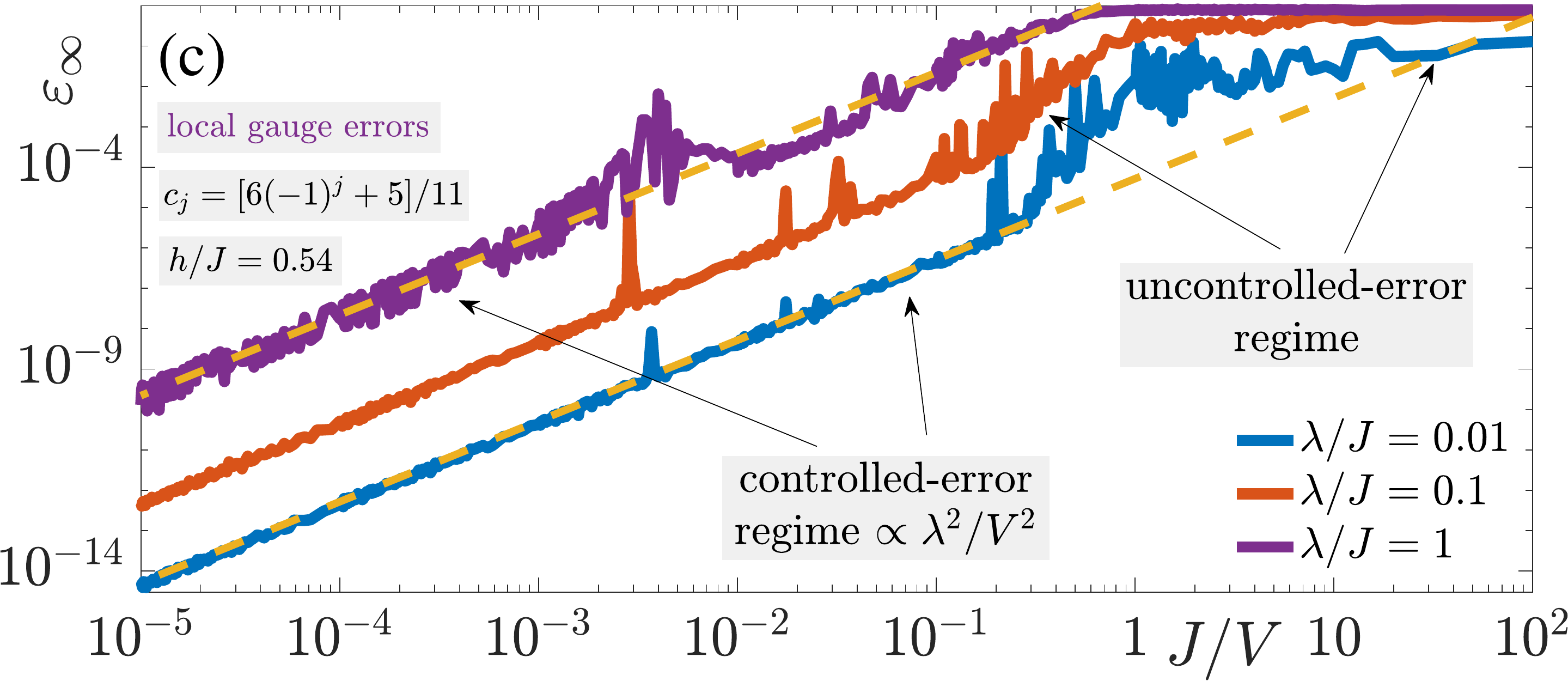}
	\caption{(Color online). Same as Fig.~\ref{fig:local}(b) but for (a) different initial states, (b) different model-parameter values, and (c) different error strengths. In all cases, the qualitative picture of two regimes persists, where at values of $V$ that are too small the long-time violation cannot be directly determined from the value of $V$, but whereas at sufficiently large $V$ the long-time violation enters a regime of controlled error $\propto\lambda^2/V^2$.}
	\label{fig:diff} 
\end{figure}

In the main text, we have focused on the staggered-matter initial state $\ket{\psi_0}$ shown in Fig.~\ref{fig:InitialStates}. However, LPG protection works for generic initial states within a gauge-invariant sector. In keeping with experimental relevance, we quench different initial product states $\ket{\psi_0'}$ and $\ket{\psi_0''}$, shown in Fig.~\ref{fig:InitialStates}, with the faulty Hamiltonian $\hat{H}=\hat{H}_0+\lambda \hat{H}_1+V\sum_j\hat{W}_j[6(-1)^j+5]/11$. We look at the long-time gauge violation as a function of $J/V$, which is displayed in Fig.~\ref{fig:diff}(a). The conclusion is qualitatively and, more or less, quantitatively the same between the three considered initial states, with a clear transition from an uncontrolled-error regime at small enough $V$, to a controlled-error $\propto\lambda^2/V^2$ regime at sufficiently large $V$.

This robustness to initial conditions is also present when it comes to different values of the model parameters. Fixing $\lambda/J=0.01$, and quenching $\ket{\psi_0}$ with $\hat{H}=\hat{H}_0+\lambda \hat{H}_1+V\sum_j\hat{W}_j[6(-1)^j+5]/11$, we find that the long-time violation exhibits the same qualitative transition between uncontrolled and controlled error $\propto\lambda^2/V^2$ as a function of $J/V$ regardless of the value of $h/J$, as shown in Fig.~\ref{fig:diff}(b). Note that the $(1+1)-$D $\mathbb{Z}_2$ LGT has a phase transition from a deconfined phase at $h/J=0$ to a confined phase at $h/J>0$ \cite{Borla2019}, but LPG protection works efficiently in either phase, at least for the system sizes considered.

In the main text, we have focused on perturbative errors ($\lambda/J<1$), but LPG protection works also for nonperturbative errors, as demonstrated in Fig.~\ref{fig:diff}(c). Here we again quench $\ket{\psi_0}$ with $\hat{H}=\hat{H}_0+\lambda \hat{H}_1+V\sum_j\hat{W}_j[6(-1)^j+5]/11$, and plot the infinite-time gauge violation as a function of $J/V$ for various values of $\lambda/J$, including the nonperturbative regime $\lambda=J$. The qualitative behavior of a transition between an uncontrolled-error regime for small enough $V$ to one with a controlled violation $\propto\lambda^2/V^2$ at sufficiently large $V$ persists regardless of the value of $\lambda$. Naturally, the larger $\lambda$ is, the larger the value of the minimal $V$ required to be in the controlled-error regime. However, we note that typical error strengths in modern QSM setups are usually $\lambda/J<1$ \cite{Schweizer2019}.

\begin{figure}[t!]
	\centering
	\includegraphics[width=.48\textwidth]{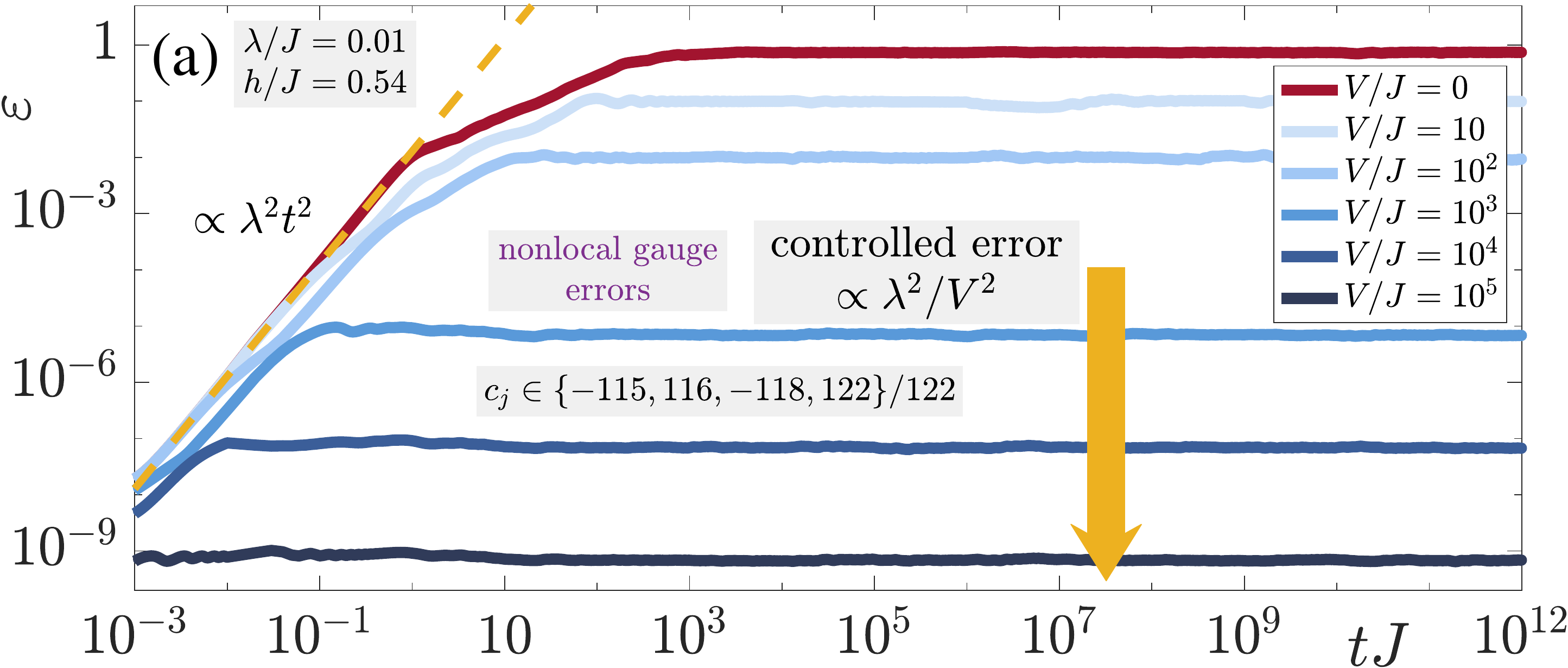}\\
	\vspace{1.1mm}
	\includegraphics[width=.48\textwidth]{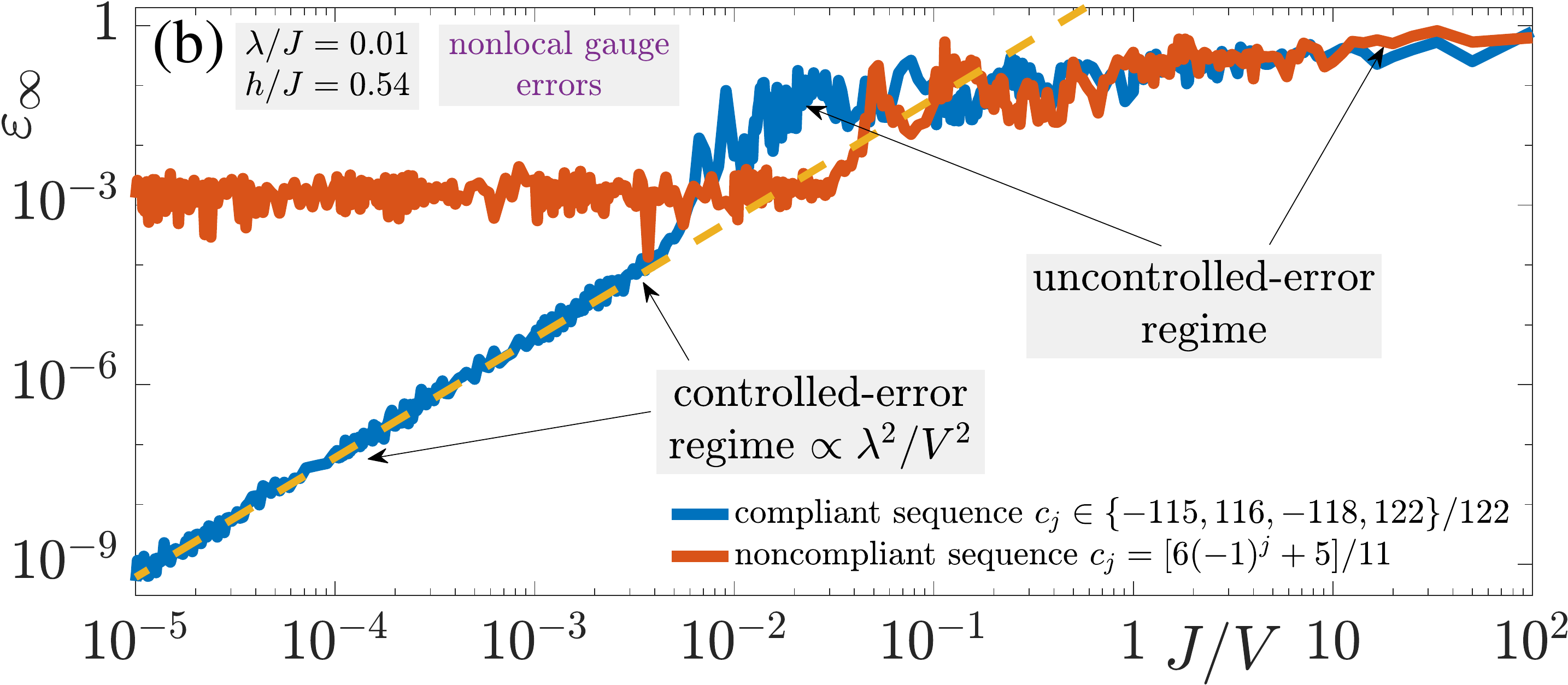}\\
	\vspace{1.1mm}
	\includegraphics[width=.48\textwidth]{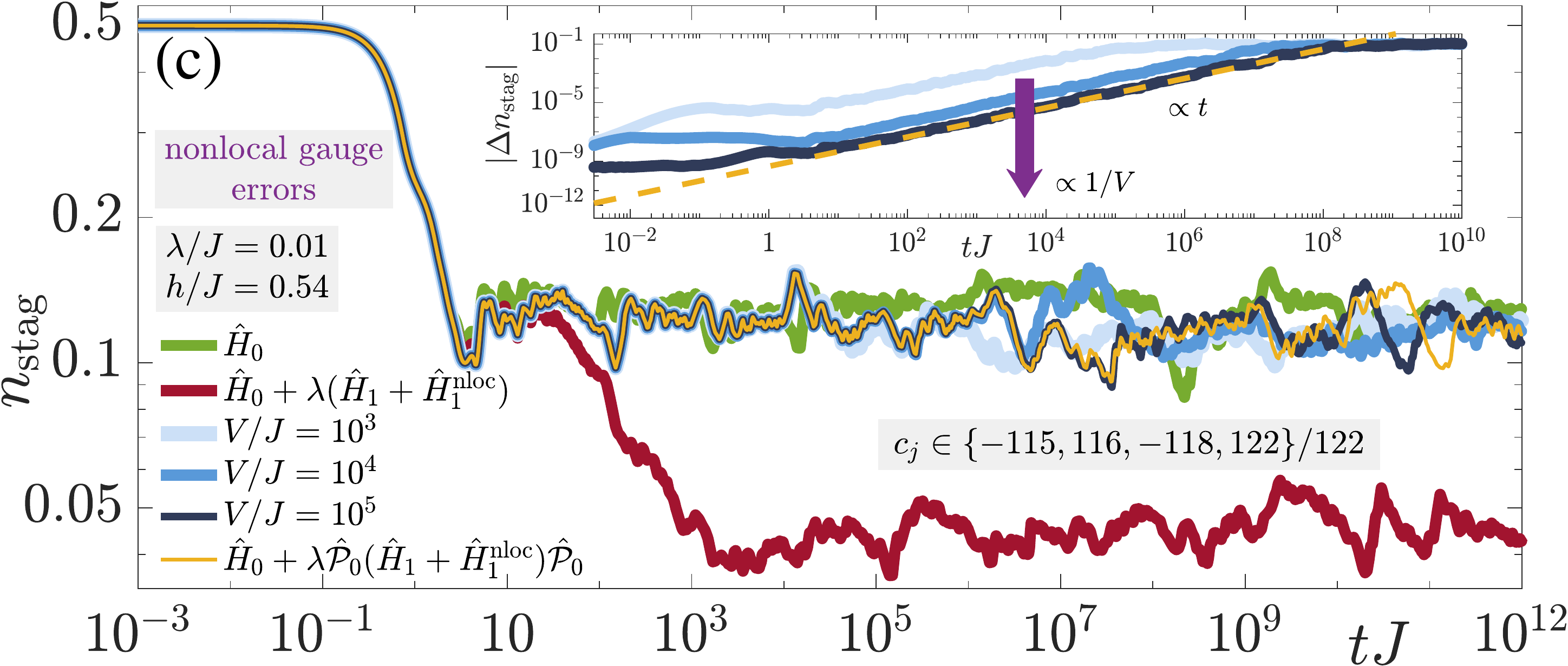}
	\caption{(Color online). Same as Fig.~\ref{fig:nonlocal} but the nonlocal gauge-breaking error now reads $\hat{H}_1^\text{nloc}=\sum_{\xi=\pm1}\prod_j\big[\mathds{1}+\xi\hat{\tau}^z_{j,j+1}\big]\big[\mathds{1}+\xi\big(\hat{a}_j+ \hat{a}^\dagger_j\big)\big]$, which also violates boson-number conservation. Here we restrict to a system of only $L=4$ matter sites to reduce numerical overhead in light of the large evolution times we access. The initial state is the staggered-matter product state $\ket{\psi_0}$ of Fig.~\ref{fig:InitialStates} but with only $L=4$ matter sites and $L=4$ gauge links. The qualitative picture is identical to that of Fig.~\ref{fig:nonlocal}.}
	\label{fig:fullBasis} 
\end{figure}
\subsection{Results with gauge-breaking errors that do not conserve boson number}\label{app:number}
In the main text, we have rigorously tested the LPG protection with a compliant sequence against local and nonlocal errors, where both conserve boson number, as does the ideal gauge theory $\hat{H}_0$. Even though $\hat{H}_1$ naturally hosts a global $\mathrm{U}(1)$ symmetry as derived in Ref.~\cite{Schweizer2019}, we have chosen $\hat{H}_1^\text{nloc}$ in Eq.~\eqref{eq:nonloc} to also conserve boson number in order to reduce numerical overhead and reach $L=6$ matter sites in our ED calculations within the bosonic half-filling sector for the large evolution times we access. However, our conclusions are independent of whether or not the global $\mathrm{U}(1)$ symmetry associated with boson-number conservation is preserved. We test this assertion by modifying the nonlocal gauge-breaking error into the form
\begin{align}
\hat{H}_1^\text{nloc}=\sum_{\xi=\pm1}\prod_j\big[\mathds{1}+\xi\hat{\tau}^z_{j,j+1}\big]\big[\mathds{1}+\xi\big(\hat{a}_j+ \hat{a}^\dagger_j\big)\big],
\end{align}
which restricts us numerically to $L=4$ matter sites. We quench the corresponding staggered-matter initial state $\ket{\psi_0}$ by the faulty theory $\hat{H}=\hat{H}_0+\lambda (\hat{H}_1+\hat{H}_1^\text{nloc})+V\sum_jc_j\hat{W}_j$ with the compliant sequence $c_j\in\{-115,116,-118,122\}/122$. Even though we set $\lambda/J=0.01$ and $h/J=0.54$, we have checked that our results hold for other values of these parameters. In Fig.~\ref{fig:fullBasis}(a), we show the ensuing dynamics of the gauge violation. The qualitative behavior is identical to the case of boson-number-conserving nonlocal errors discussed in the main text. The gauge violation grows initially $\propto\lambda^2t^2$, in agreement with time-dependent perturbation theory \cite{Halimeh2020a}, before settling into a plateau. The latter shows no direct relation to the protection strength when $V$ is too small. However, at sufficiently large $V$, the violation plateau begins at a timescale $\propto1/V$ and takes on a value $\propto\lambda^2/V^2$. This behavior is further confirmed in Fig.~\ref{fig:fullBasis}(b), which shows the long-time violation as a function of $J/V$. At values of $V$ that are too small, the violation is uncontrolled, whereas at sufficiently large $V$, the long-time violation is controlled and scales $\propto\lambda^2/V^2$. Note once again how the noncompliant sequence $c_j=[6(-1)^j+5]/11$ is not sufficient to achieve reliable gauge invariance in the case of nonlocal errors here, regardless of how large $V$ is. Instead, it seems to always remain above a certain minimal value.

The dynamics of the staggered boson number, Eq.~\eqref{eq:nstag}, is shown in Fig.~\ref{fig:fullBasis}(c), and the qualitative picture is the same as that of a nonlocal error that conserves boson number, see Fig.~\ref{fig:nonlocal}(c). Indeed, we find that an adjusted gauge theory $\hat{H}_\text{adj}=\hat{H}_0+\lambda\hat{\mathcal{P}}_0(\hat{H}_1+\hat{H}_1^\text{nloc})\hat{\mathcal{P}}_0$ faithfully reproduces the dynamics of the local observable up to a timescale $\tau_\text{adj}\propto V/(V_0L)^2$, with an error that is suppressed $\propto1/V$ and grows linearly in time (see inset of Fig.~\ref{fig:fullBasis}(c)), as predicted analytically (see Sec.~\ref{sec:adjusted}).

\subsection{Local-error coefficients $\alpha_m$}\label{sec:coeffs}
The coefficients $\alpha_{1\ldots4}$ of the local error term, Eq.~\eqref{eq:H1}, in the $1$D $\mathbb{Z}_2$ LGT are inspired from an extended version of building-block errors arising in the construction of the effective Floquet Hamiltonian in the experiment of Ref.~\cite{Schweizer2019}. Explicitly, they read
\begin{subequations}
	\begin{align}\nonumber
	\alpha_1=&\,\sum_{k>0}\frac{\mathcal{K}(\chi)}{k}\big[\mathcal{J}_{-k-1}(\chi)\mathcal{J}_{-k-2}(\chi)+\mathcal{J}_k(\chi)\mathcal{J}_{k+1}(\chi)\\
	&-\mathcal{J}_{k-1}(\chi)\mathcal{J}_{k-2}(\chi)-\mathcal{J}_{-k}(\chi)\mathcal{J}_{-k+1}(\chi)\big],\\\nonumber
	\alpha_2=&\,\sum_{k>0}\frac{\mathcal{K}(\chi)}{k}\big[\mathcal{J}_{-k+1}(\chi)\mathcal{J}_{k-2}(\chi)+\mathcal{J}_{-k}(\chi)\mathcal{J}_{k-1}(\chi)\\
	&-\mathcal{J}_{k+1}(\chi)\mathcal{J}_{-k-2}(\chi)-\mathcal{J}_{k}(\chi)\mathcal{J}_{-k-1}(\chi)\big],\\\nonumber
	\alpha_3=&\,\sum_{k>0}\frac{\mathcal{K}(\chi)}{k}\big[\mathcal{J}_{k-1}^2(\chi)+\mathcal{J}_{k-2}^2(\chi)\\
	&-\mathcal{J}_{-k-1}^2(\chi)-\mathcal{J}_{-k-2}^2(\chi)\big],\\\nonumber
	\alpha_4=&\,\sum_{k>0}\frac{\mathcal{K}(\chi)}{k}\big[\mathcal{J}_{-k+1}^2(\chi)+\mathcal{J}_{-k}^2(\chi)\\
	&-\mathcal{J}_{k+1}^2(\chi)-\mathcal{J}_{k}^2(\chi)\big],
	\end{align}
\end{subequations}
where $\mathcal{J}_q(\chi)$ is the Bessel function of the first kind and order $q$, and the variable $\chi$ is a dimensionless driving parameter that is set to the experimentally relevant \cite{Schweizer2019} value $\chi=1.84$ for the related results of this work, although we have checked that our qualitative picture is independent of the choice of $\chi$. We have also used $\mathcal{K}(\chi)$ as a nonzero factor enforcing $\sum_{m=1}^4\alpha_m=1$, in order to encapsulate the overall strength of the error terms solely in $\lambda$.

\section{Emergent gauge theories}\label{sec:analytics}
In this Appendix, we detail the analytic derivations for the emergent gauge theories arising under LPG protection, along with the associated timescales appearing in our numerical results.

\subsection{Renormalized gauge theory}\label{sec:renormalized}
One can adapt techniques employed by Abanin, De Roeck, Ho, and Huveneers (ARHH) \cite{abanin2017rigorous} for slow heating in periodically driven systems to show how the LPG protection with a compliant sequence can stabilize gauge invariance up to a timescale exponential in $V$, thereby explaining the remarkable stability of gauge invariance present up to all accessible evolution times in our numerical results. This adaptation has already been done for the quadratic energy-penalty protection in the case of Abelian \cite{vandamme2021reliability} and also non-Abelian \cite{Halimeh2021gauge} LGTs, and also for the protection linear in the full generator $\hat{G}_j$ with a compliant sequence in the case of the Abelian $\mathrm{U}(1)$ LGT \cite{Halimeh2020e}.

Let us define $\Lambda$ as a finite subset of a $d$-dimensional cubic lattice, and $\mathcal{B}_\Lambda$ the algebra of bounded operators acting within the corresponding total Hilbert space $\mathcal{H}_\Lambda$. We consider the operators $\hat{O}_S\otimes\hat{\mathds{1}}_{\Lambda\backslash S}$ acting within the subset $S\subset\Lambda$ with a subalgebra $\mathcal{B}_S\in\mathcal{B}_\Lambda$ equipped with the standard operator norm. An operator $\hat{O}$ can now be nonuniquely decomposed as 
\begin{align}
\hat{O}=\sum_{S\in\Pi_c(\Lambda)}\hat{O}_S,
\end{align}
with the \textit{interaction potential} $\hat{O}_S\in\mathcal{B}_S$, and where $\Pi_c(\Lambda)$ is the set of finite subsets of $\Lambda$ that are connected by adjacency (i.e., adjacent sites). Let us define a family of norms on potentials parametrized by a rate $\kappa>0$ that allows for different weights to be assigned to operators with different spatial support:
\begin{align}
\lvert\lvert\hat{X}\rvert\rvert_\kappa \coloneqq \sup\limits_{x\in\Lambda} \sum_{S\in \Pi_c(\Lambda),\, x\in S} e^{\kappa |S|} \lvert\lvert\hat{X}_S\rvert\rvert.
\end{align}
Here, the supremum is found on the lattice site $x$ where the operators $\hat{X}_S$ with finite support on $x$ yield the largest sum in their weighted norms.

Let us call $\hat{\mathcal{P}}_m$ the projector onto the eigenstates with eigenvalue $m$ of the LPG protection Hamiltonian $\hat{H}_\text{pro}$, where 
\begin{align}\label{eq:Hpro}
\tilde{V}\hat{H}_\text{pro}=V\hat{H}_W=V\sum_jc_j\big[\hat{W}_j(g_j^\text{tar})-g_j^\text{tar}\big],
\end{align}
$c_j$ is a rational compliant sequence, i.e., it satisfies $\sum_jc_j[w_j(g_j^\text{tar})-g_j^\text{tar}]=0\iff w_j(g_j^\text{tar})=g_j^\text{tar},\,\forall j$, and $\tilde{V}$ is such that the spectrum of $\hat{H}_\text{pro}$ is composed of integers ($m\in\mathds{Z}$). The latter is not possible to have if $c_j$ are not rational. Note that $\hat{\mathcal{P}}_0$ is the projector onto the target gauge sector $g_j=g_j^\text{tar},\,\forall j$. The faulty gauge theory $\hat{H}=\hat{H}_0+\lambda \hat{H}_1+V\hat{H}_W$ can then be split into two parts: $\hat{H}_\text{d}+V\hat{H}_W$, with \begin{align}
\hat{H}_\text{d}=\sum_m\hat{\mathcal{P}}_m \big(\hat{H}_0+\lambda \hat{H}_1\big)\hat{\mathcal{P}}_m,
\end{align}
which is invariant with respect to $\hat{H}_\text{pro}$, and the remaining term is
\begin{align}
\hat{H}_\text{nd}=\hat{H}-\hat{H}_\text{d}-V\hat{H}_W.
\end{align}
As such, $[\hat{H}_\text{d},\hat{H}_W]=0$, although $\hat{H}_\text{d}$ does not necessarily commute with either $\hat{G}_j$ or $\hat{W}_j$, and we also know that by construction $[\hat{H},\hat{H}_W]\neq0$. Indeed, $\hat{H}_\text{d}$ and $\hat{H}_W$ share the same global symmetry generated by the latter, but not the local pseudo gauge symmetry whose generator is $\hat{W}_j$ or the local full gauge symmetry whose generator is $\hat{G}_j$.

In following with the ARHH framework \cite{abanin2017rigorous}, let us assume that there exist a rate $\kappa_0$ leading to the relevant energy scale
\begin{align}
V_0\coloneqq\frac{54\pi}{\kappa_0^2}\big(\lvert\lvert \hat{H}_\text{d}\rvert\rvert_{\kappa_0}+2\lvert\lvert \hat{H}_\text{nd}\rvert\rvert_{\kappa_0}\big),
\end{align}
and that, in addition to the compliance condition, we fulfill the following two conditions:
\begin{subequations}\label{eq:conditions}
\begin{align}
&\tilde{V}\geq\frac{9\pi}{\kappa_0}\lvert\lvert \hat{H}_\text{nd}\rvert\rvert_{\kappa_0},\\
&\big\lfloor V_0^{-1}(1-\ln{V_0}+\ln{\tilde{V}})^{-3}\tilde{V}\big\rfloor - 3 \ge 0. 
\end{align}
\end{subequations}
Once these conditions are satisfied, then starting in any initial state $\ket{\psi_0}$ within the target gauge sector $g_j=g_j^\text{tar},\,\forall j$, will give rise to dynamics where the gauge violation remains bounded from above as
\begin{align}
\big\lvert\bra{\psi_0}e^{i\hat{H}t}\hat{\mathcal{G}}e^{-i\hat{H}t}\ket{\psi_0}\big\rvert<\frac{K(\hat{\mathcal{G}})}{\tilde{V}},
\end{align}
up to a timescale $\tau_\text{ren}\propto V_0^{-1}e^{\tilde{V}/V_0}$, where $\hat{\mathcal{G}}=\sum_j(\hat{G}_j-g_j^\text{tar})^2/L$ is the gauge-violation operator, and $K$ is a model-parameter-dependent term, but which is independent of $\tilde{V}$ and system size. 

Details of this proof in the context of gauge protection have been outlined in Ref.~\cite{Halimeh2020e}. The latter work deals specifically with a protection term linear in the full generator $\hat{G}_j$ with a rational compliant sequence. However, since the LPG protection $\tilde{V}\hat{H}_\text{pro}=V\hat{H}_W$ satisfies the condition of compliance, and as Eqs.~\eqref{eq:conditions} are also satisfied, the derivation of Ref.~\cite{Halimeh2020e} applies in full also here, and, as such, we refer the interested reader there for its details.

Nevertheless, a few comments are in order. Even though the timescale $\tau_\text{ren}\propto V_0^{-1}e^{\tilde{V}/V_0}$ may not appear directly volume-dependent, a larger $V$ is required with larger system size $L$ in order to achieve a given level of reliability. This becomes clear when looking at Eq.~\eqref{eq:Hpro}. As mentioned, $c_j$ form a compliant sequence of rational numbers normalized such that $\max_j\{\lvert c_j\rvert\}=1$. Let us call $f_j$ the set of smallest integers such that $f_j/\max_m\{\lvert f_m\rvert\}=c_j$. As such, we can rewrite 
\begin{align}
\tilde{V}\hat{H}_\text{pro}=\frac{V}{\max\limits_j\{\lvert f_j\rvert\}}\sum_jf_j\big[\hat{W}_j(g_j^\text{tar})-g_j^\text{tar}\big],
\end{align}
meaning that $\tilde{V}=V/\max_j\{\lvert f_j\rvert\}$ is sufficient to make the spectrum of $\hat{H}_\text{pro}$ integer. Assuming that a given value of $\tilde{V}$ brings about a certain level of gauge-error suppression, a larger system size will lead to a larger $\max_j\{\lvert f_j\rvert\}$, meaning that $V$ has to become larger in order to retain the same value of $\tilde{V}$. Naturally, this becomes intractable in the thermodynamic limit. However, we also see in our ED calculations that even the noncompliant sequence, which does not grow with system size, achieves reliable protection for local errors up to indefinite times, even though we cannot analytically predict this. The nonlocal errors we have considered in this work are very drastic, and only such errors require the compliant sequence.

Another point worth mentioning is that our analytic arguments for the compliant sequence strictly only apply for local errors, and extreme nonlocal errors with support over the whole lattice in the thermodynamic limit are not within the operator algebras we have defined. However, as we see in our numerical results, LPG protection with a compliant sequence still suppresses gauge violations up to indefinite times even in the presence of such extreme errors on a finite system, and this is within the ARHH framework but cannot be guaranteed in the thermodynamic limit. Furthermore, LPG protection with a noncompliant sequence, which does not fulfill all the conditions of the ARHH formalism, still offers stable gauge invariance up to indefinite times when gauge-breaking errors are local. This cannot be guaranteed by the ARHH framework, but it is not ruled out either. Indeed, this formalism gives a guaranteed minimal (worst-case scenario) timescale exponential in $V$ up to which gauge invariance is stabilized in the presence of errors with a finite spatial support (that does not grow with system size) given that the compliance condition and Eqs.~\eqref{eq:conditions} are satisfied, but it does not forbid stable gauge invariance when any of these conditions are not strictly met.

Finally, it is to be noted that obtaining a closed form of the renormalized gauge theory is generically difficult. Moreover, we cannot numerically test how faithfully such a renormalized gauge theory reproduces the LPG-protected dynamics under the faulty theory, as this would require reaching exponentially long times within systems in the thermodynamic limit, for which no general techniques exist.

\subsection{Adjusted gauge theory}\label{sec:adjusted}
It is useful for ongoing experiments to be able to have an exact form of an emergent gauge theory in the wake of a quench with the faulty gauge theory $\hat{H}=\hat{H}_0+\lambda \hat{H}_\text{err}+V\hat{H}_W$, where $\hat{H}_\text{err}=\hat{H}_1+\eta \hat{H}_1^\text{nloc}$ with $\eta=0$ or $1$. One can show through the quantum Zeno effect (QZE) \cite{facchi2002quantum,facchi2004unification,facchi2009quantum,burgarth2019generalized} in the case of LPG protection with a compliant or suitably chosen noncompliant sequence at sufficiently large protection strength $V$, that an adjusted gauge theory $\hat{H}_\text{adj}=\hat{H}_0+\lambda \hat{\mathcal{P}}_0\hat{H}_\text{err}\hat{\mathcal{P}}_0$ arises up to a timescale $\tau_\text{adj}\propto V/(V_0L)^2$ \cite{Halimeh2020e}. Specifically, at sufficiently large $V$ the dynamics under $\hat{H}$ is restricted to the ``decoherence-free'' subspace of $\hat{H}_W$. In the large-$V$ limit, the time-evolution operator reads \cite{facchi2002quantum,facchi2004unification,facchi2009quantum,burgarth2019generalized}
\begin{align}\label{eq:largeV}
\lim_{V\to\infty}e^{-i\hat{H}t}=e^{-i[V\hat{H}_W+\sum_m\hat{\mathcal{P}}_m(\hat{H}_0+\lambda \hat{H}_\text{err})\hat{\mathcal{P}}_m]t},
\end{align}
up to a residual additive term $\propto V_0^2L^2t/V$. We now consider the conditions for which the QZE can promise reliable stabilization of gauge invariance in the dynamics up to the resulting timescale $\tau_\text{adj}\propto V/(V_0L)^2$.

\subsubsection{$\hat{H}_W$ is nondegenerate}
In this case, gauge invariance is stable for a generic $\hat{H}_\text{err}$ so long as the coefficients $c_j$ are \textit{sufficiently incommensurate}. In other words, given any two \textit{pseudo superselection sectors} $\mathbf{w}=(w_1,w_2,\ldots)$ and $\mathbf{w}'=(w_1',w_2',\ldots)$ of $\hat{W}_j,\,\forall j$, then the sequence must satisfy $\sum_jc_j(w_j-w_j')\neq0$. This condition is readily satisfied when $c_j$ is a sequence of random or irrational numbers, for example. 

We note here that a pseudo superselection sector $\mathbf{w}$ of $\hat{H}_W$ is not necessarily gauge-invariant except when it coincides with the target sector, i.e., when $\mathbf{w}=\mathbf{g}^\text{tar}=(g_1^\text{tar},g_2^\text{tar},\ldots)$.

\subsubsection{$\hat{H}_W$ is degenerate}
In the case the term $\hat{H}_0+\lambda \hat{H}_\text{err}$ does not lift the degeneracy of $\hat{H}_W$ in first-order perturbation theory, then we can utilize that
\begin{align}\nonumber
\hat{\mathcal{P}}_m\big(\hat{H}_0+\lambda \hat{H}_\text{err}\big)\hat{\mathcal{P}}_m&=\sum_{\mathbf{w},\mathbf{w}'\in\mathcal{D}_m}\hat{P}_\mathbf{w}\big(\hat{H}_0+\lambda \hat{H}_\text{err}\big)\hat{P}_{\mathbf{w}'}\\
&=\sum_{\mathbf{w}\in\mathcal{D}_m}\hat{P}_\mathbf{w}\big(\hat{H}_0+\lambda \hat{H}_\text{err}\big)\hat{P}_{\mathbf{w}},
\end{align}
where $\mathcal{D}_m$ is the set of all pseudo superselection sectors $\mathbf{w}$ of $\hat{W}_j$ such that $\hat{H}_W\ket{\psi}=m\ket{\psi},\,\forall\ket{\psi}\in\mathbf{w}$, and $\hat{P}_\mathbf{w}$ is the projector onto the pseudo superselection sector $\mathbf{w}$. Consequently, Eq.~\eqref{eq:largeV} becomes

\begin{align}\label{eq:largeVnext}
\lim_{V\to\infty}e^{-i\hat{H}t}=e^{-i\sum_m\sum_\mathbf{w\in\mathcal{D}_m}[mV\hat{P}_\mathbf{w}+\hat{P}_\mathbf{w}(\hat{H}_0+\lambda \hat{H}_\text{err})\hat{P}_\mathbf{w}]t}.
\end{align}
Since we prepare our initial state in the target sector $\mathbf{w}=(g_1^\text{tar},g_2^\text{tar},\ldots)$, gauge-noninvariant processes driving the dynamics out of this sector will be suppressed in the time evolution for large $V$, because different sectors do not couple in the QZE regime as evidenced in Eq.~\eqref{eq:largeVnext}, and this is precisely because second-order perturbation theory is beyond the timescale of QZE protection.

As mentioned, LPG protection can be shown to stabilize gauge invariance for an adequately chosen, yet not necessarily compliant, sequence $c_j$ through an effective QZE behavior up to a residual additive term $\propto t(V_0L)^2/V$. In particular, the latter can be formulated as
\begin{align}\nonumber
&\big\lvert\big\lvert e^{-i\hat{H}t}-e^{-i[V\hat{H}_W+\sum_m\hat{\mathcal{P}}_m(\hat{H}_0+\lambda \hat{H}_\text{err})\hat{\mathcal{P}}_m]t}\big\rvert\big\rvert\\
&\leq\mathcal{Q}\propto\frac{tV_0^2L^2}{V}.
\end{align}
Projecting onto the target sector, this becomes
\begin{align}\nonumber
&\big\lvert\big\lvert \hat{\mathcal{P}}_0\big[e^{-i\hat{H}t}-e^{-i(\hat{H}_0+\lambda\hat{\mathcal{P}}_0\hat{H}_\text{err}\hat{\mathcal{P}}_0)t}\big]\hat{\mathcal{P}}_0\big\rvert\big\rvert\\\label{eq:QZEinequality}
&\leq\mathcal{Q}\propto\frac{tV_0^2L^2}{V},
\end{align}
where here we have utilized the fact that in the target sector, where we initialize our system, $\hat{H}_0$ and $\hat{\mathcal{P}}_0\hat{H}_0\hat{\mathcal{P}}_0$ drive identical dynamics, and so the adjusted gauge theory $\hat{H}_\text{adj}=\hat{H}_0+\lambda \hat{\mathcal{P}}_0\hat{H}_\text{err}\hat{\mathcal{P}}_0$ has naturally appeared in our formalism. It is to be noted, however, that the adjusted gauge theory can also be derived through the formalism of constrained quantum dynamics in the case of full protection \cite{gong2020universal,gong2020error}. 

As we will show in the following, the inequality~\eqref{eq:QZEinequality} translates to the dynamics of a local observable $\hat{O}$ under the faulty theory being gauge-invariant up to an error upper bound $\propto t(V_0L)^2/V$. The dynamics of a local observable $\hat{O}$ under the faulty theory $\hat{H}$ deviates from that under the adjusted gauge theory as
\begin{align}\nonumber
&\big\lvert \bra{\psi(t)}e^{i\hat{H}t}\hat{O}e^{-i\hat{H}t}-e^{i\hat{H}_\text{adj}t}\hat{O}e^{-i\hat{H}_\text{adj}t}\ket{\psi(t)}\big\rvert\\\nonumber
\leq&\big\lvert\big\lvert \hat{\mathcal{P}}_0\big(e^{i\hat{H}t}\hat{O}e^{-i\hat{H}t}-e^{i\hat{H}_\text{adj}t}\hat{O}e^{-i\hat{H}_\text{adj}t}\big)\hat{\mathcal{P}}_0\big\rvert\big\rvert\\\nonumber
=&\frac{1}{2}\big\lvert\big\lvert \hat{\mathcal{P}}_0\big\{\big(e^{i\hat{H}t}-e^{i\hat{H}_\text{adj}t}\big)\hat{O}e^{-i\hat{H}t}\\\nonumber
&+e^{i\hat{H}t}\hat{O}\big(e^{-i\hat{H}t}-e^{-i\hat{H}_\text{adj}t}\big)\\\nonumber
&+\big(e^{i\hat{H}t}-e^{i\hat{H}_\text{adj}t}\big)\hat{O}e^{-i\hat{H}_\text{adj}t}\\\nonumber
&+e^{i\hat{H}_\text{adj}t}\hat{O}\big(e^{-i\hat{H}t}-e^{-i\hat{H}_\text{adj}t}\big)\big\}\hat{\mathcal{P}}_0\big\rvert\big\rvert\\
\leq&\tilde{\mathcal{Q}}\propto 2\frac{tV_0^2L^2}{V}.
\end{align}
As such, we have proven that the adjusted gauge theory $\hat{H}_\text{adj}$ faithfully reproduces the dynamics of a local observable $\hat{O}$ under the faulty theory $\hat{H}$ with large $V$ up to a timescale $\tau_\text{adj}\propto V/(V_0L)^2$. This is very promising for ongoing QSM setups implementing LGTs, since it means that an emergent exact gauge theory can still be derived in closed form and realized experimentally, allowing for a controlled assessment of the fidelity of the realization.

\bibliography{Z2protection_biblio}
\end{document}